# Phase-field simulation of liquid-vapor equilibrium and evaporation of fluid mixtures

*Olivier J.J. Ronsin* [a,b]*, *DongJu Jang*[c], *Hans-Joachim Egelhaaf*[c, d, e], *Christoph J. Brabec*[d,e] *and Jens Harting* [a,b,f]*

[a] Forschungszentrum Jülich GmbH, Helmholtz Institute Erlangen-Nürnberg for Renewable Energy (IEK-11), Fürther Straße 248, 90429 Nürnberg, Germany

[b] Department of Chemical and Biological Engineering, Friedrich-Alexander-Universität Erlangen-Nürnberg, Fürther Straße 248, 90429 Nürnberg, Germany

[c] ZAE Bayern—Solar Factory of the Future, Energy Campus Nürnberg, Fürther Straße 250, 90429 Nürnberg, Germany

[d] Institute of Materials for Electronics and Energy Technology (i-MEET), Friedrich-Alexander-Universität Erlangen-Nürnberg, 91058 Erlangen, Germany

[e] Forschungszentrum Jülich GmbH, Helmholtz Institute Erlangen-Nürnberg for Renewable Energy (IEK-11), Immerwahrstrasse 2, 91058 Erlangen, Germany

[f] Department of Physics, Friedrich-Alexander-Universität Erlangen-Nürnberg, Fürther Straße 248, 90429 Nürnberg, Germany

## <u>Keywords</u>

## <u>Abstract</u>

In solution-processing of thin films, the material layer is deposited from a solution composed of several solutes and solvents. The final morphology and hence the properties of the film often depend on the time needed for the evaporation of the solvents. This is typically the case for organic photoactive or electronic layers. Therefore, it is important to be able to predict the evaporation kinetics of such mixtures. We propose here a new phase-field model for the simulation of evaporating fluid mixtures and simulate their evaporation kinetics. Similar to the Hertz-Knudsen theory, the local liquid-vapor equilibrium is assumed to be reached at the film surface and evaporation is driven by diffusion away from this gas layer. In the situation where the evaporation is purely driven by the liquid-vapor equilibrium, the simulations match the behavior expected theoretically from the free energy: for evaporation of pure solvents, the evaporation rate is constant and proportional to the vapor pressure. For mixtures, the evaporation rate is in general strongly time-dependent because of the changing composition of the film. Nevertheless, for highly non-ideal mixtures, such as poorly compatible fluids or polymer solutions, the evaporation rate becomes almost constant in the limit of low Biot numbers. The results of the simulation have been successfully compared to experiments on a polystyrene-toluene mixture. The model allows to take into account deformations of the liquid-vapor interface and therefore to simulate film roughness or dewetting.

# Introduction

Solution deposition of thin films is a widespread and crucial processing route for applications such as membrane technologies, organic electronics[1], third-generation photovoltaics[2] including organic or perovskite solar cells[3], batteries[4] and hydrogen-based energy storage and conversion systems [5]. In many of these applications, one or more solutes are dissolved in a single solvent or a mixture of solvents and the film morphology forms during the fabrication process. Depending on the material properties, various desired or undesired processes such as chemical reactions, crystallization, liquid-liquid phase-separation by spinodal decomposition or nucleation and growth can take place. A large part of these transformations occurs during the drying phase because the presence of the solvents allows for fast kinetics. As a consequence, the drying phase is crucial for the film morphology and hence the final film properties and device functionality [6][7].

For example, in organic photoactive layers made of an electron donor and an electron acceptor material, it is desired to obtain a so-called "bulk-heterojunction" structure, whereby both materials form interpenetrated, well separated phases at the nanoscale. This ensures good exciton generation within the phases and efficient separation into free charge carriers at the interfaces. Moreover, crystalline phases of both materials should be available to ensure charge extraction [8][9]. Such a morphology is crucial to obtain performant devices and has successfully contributed to the improvement of power conversion efficiency to up to 16-18% in the last decade [10][11].

It has been therefore recognized that the solvent or solvent blend used for deposition is a tool of choice to gain control over the formation of the morphology [12][13]. On the one hand, the thermodynamic properties of the mixture can be modulated in order to trigger or hinder phase transitions. This is typically the case for liquid-liquid phase separation. On the other hand, the evaporation kinetics is determined by the solvents, and the time available for fast structure formation before the film is kinetically "quenched" in the solid state can therefore be adjusted by changing the solvents.

In order to better understand the process-structure relationship in solution-processed thin films, tremendous efforts have been made to simulate the drying stage of phase-separating or crystallizing systems, with various simulation methods from the molecular scale [14][15] to the continuum level [16][17][18][19][20][21]. Nevertheless, these works focus on the (evaporation-induced) phase transformations in the film, while the description of the evaporation process itself remains rather simple. This may be improved so that the simulated time-dependent concentrations and drying times better match the experimental evidence, which should result in more accurate predictions of the final film morphology.

Significant progress has also been made in continuum simulations of boiling and evaporation with lattice Boltzmann models[22] as well as with phase-field or coupled phase-field Navier-Stokes methods [23][24][25]. However, these models do not fulfill all requirements needed in order to use them in simulations of film structure formation, especially simplicity, computational efficiency, the handling of multicomponent mixtures and of the film surface deformation. Therefore, we recently proposed a simple and efficient phase-field model for the evaporation of fluid mixtures, rendering the evaporation kinetics properly and usable in heavy simulations of evaporation-induced film structure formation [26]. However, this model fails at matching all the physics of the evaporation process. In the current work, we propose a still very simple, improved model enabling us to simulate accurately the evaporation rate of pure solvents, but also the evaporation kinetics of mixtures.

The basic representation of the evaporation process is given by the so-called Hertz-Knudsen theory [27][28]: it is assumed that a gas layer being in thermodynamic equilibrium with the liquid forms directly above the liquid surface. This layer, whose composition is given by the liquid-vapor equilibrium, contains more solvent molecules than the surrounding environment. Thus, solvent molecules diffuse to the environment and the evaporation mass flux $j$ of a pure solvent is given by

$$j = \sqrt{\frac{M}{2\pi kT}} \left( \alpha_e P_{sat} - \alpha_c P_{vap} \right) \tag{1}$$

where $M$ is the molar mass, $k$ the Boltzmann constant, $T$ the temperature, $P_{sat}$ the vapor pressure, and $P_{vap}$ the partial pressure in the environment. $\alpha_e$ and $\alpha_c$ are the evaporation and condensation coefficients, respectively. This relationship has turned out to be very successful for the prediction of solvent evaporation rates, despite some inconsistencies with experimental results that led to successive refinements of the theory over the years [29].

In the case of mixtures, the partial pressure of the gas layer close to the film, determined from the liquid-vapor equilibrium, should be used instead of $P_{sat}$ in the equation above to estimate the evaporation mass flux of each solvent. However, beyond this purely thermodynamic effect, other phenomena might play an important role in the drying kinetics of the blend, especially in the case of polymer solutions. First, the diffusion coefficients are strongly composition-dependent and may drop over several orders of magnitude upon drying [30][31][32][33][34], so that diffusion is too slow to prevent the build-up of concentration gradients close to the surface. This so-called "skin effect" leads to a strong decrease of the evaporation rate at the end of the drying [26][32][35][36][37]. Second, for polymers with a glass-transition temperature higher than the process temperature, the glassy state can be reached in the film when the amount of solvent becomes small upon drying. The drying kinetics is then limited by the relaxation properties of the polymer [30][38][39]. Third, in thin polymer films with final thicknesses that are typically in the nanometer range, the final stages of the drying kinetics can be strongly impacted by the reduced mobility of the polymer close to the substrate [39]. Fourth, the gas flow in the vapor phase strongly influences the evaporation rate, which is in practice routinely used to vary the drying kinetics with variable air flow rates [37][40]. Numerous successful one-dimensional simulations of drying, interdiffusion or sorption experiments, taking these effects into account, have been reported for various multicomponent systems [32][36][37][38][39][41].

The objective of the phase-field framework presented in the current paper is to simulate the physical situation described by the Hertz-Knudsen theory. Thereby, the liquid-vapor (LV) interface is not just a boundary condition at the top of the simulation box: it is fully included and can move inside the simulated domain, which contains not only the liquid film but also a gas phase (see **Figure 1**). In 2D or 3D, with a simple fixed regular mesh, this choice allows for simulations of film structuring upon drying that takes into account possible deformations of the surface and hence predict for example the surface roughness of the dry layer or dewetting phenomena. The present paper is about the validation of the approach, and is focused on checking that the behavior which can be expected from the thermodynamics of the drying film is properly recovered. In order to do this, we deliberately ignore all other contributions that might impact the drying kinetics in real systems (diffusion-limited drying, polymer relaxation or confinement, gas flow effects…), except for the comparison with experimental results presented at the end of the paper.

After this introduction, the model is presented in the second section. In the third section, it is compared to the results that can be expected from the theory, for pure solvents first and then for mixtures. A comparison with experiments on a drying polystyrene-toluene mixture is also presented, as well as one showcase in 2D considering drying with LV-interface deformation. This illustrates the potential of the framework for situations where the surface roughness of the final film or the coverage of the substrate are crucial.

## Model equations

### *Free energy functional*

The system to be investigated is composed of $n$ fluids which can have a liquid and a vapor phase. Its state is described by the respective volume fractions of these materials $\varphi_i$ as well as by an order parameter $\phi_{vap}$ which varies from 0 in the liquid phase to 1 in the vapor phase. Inspired by classical phase-field methods, we propose a modified version of the free energy as already used in our previous model [26]. The total free energy of the system reads

$$G_{tot} = \int_V \left( \Delta G_V^{loc} + \Delta G_V^{nonloc} \right) dV \tag{2}$$

where $V$ is the system volume. $\Delta G_V^{loc}$ is the local free energy density and $\Delta G_V^{nonloc}$ the non-local contribution due to the field gradients. The local part of the free energy is given by

$$\Delta G_V^{loc}(\{\varphi_i\}, \phi_{vap}) = \left( 1 - p(\phi_{vap}) \right) \Delta G_V^{liq}(\{\varphi_i\}) + p(\phi_{vap}) \Delta G_V^{vap}(\{\varphi_i\}) + \Delta G_V^{num}(\{\varphi_i\}) \tag{3}$$

The first term on the right-hand side of the equation above represents the free energy density change upon mixing in the liquid phase for which we use the classical Flory-Huggins theory,

$$\Delta G_V^{liq}(\{\varphi_i\}) = \frac{RT}{v_0} \left( \sum_{i=1}^n \frac{\varphi_i ln \varphi_i}{N_i} + \sum_{i=1}^n \sum_{j>i}^n \varphi_i \varphi_j \chi_{ij} \right) \tag{4}$$

with $R$ being the gas constant and $T$ the temperature. $v_0$ is the molar volume of the lattice site considered to calculate the free energy of mixing in the sense of the Flory-Huggins theory [42] $N_i$ is the molar size of the fluid $i$ in terms of units of the lattice site volume, so that its molar volume is $v_i = N_i v_0$. $\chi_{ij}$ is the interaction parameter between the amorphous phases of materials $i$ and $j$. In the gas phase, for simplicity, the mixture is assumed to be ideal with gases of the same molecular size, and the local free energy contribution reads

$$\Delta G_V^{vap}(\{\varphi_i\}) = \frac{RT}{v_0} \sum_{i=1}^n \varphi_i ln \left( \frac{P_i}{P_{sat,i}} \right) = \frac{RT}{v_0} \sum_{i=1}^n \varphi_i ln \left( \frac{\varphi_i}{\varphi_{sat,i}} \right) \tag{5}$$

where we define in the gas phase $\varphi_{sat,i} = P_{sat,i}/P_0$ and $\varphi_i = P_i/P_0$, $P_{sat,i}$ and $P_i$ being the vapor pressure of the fluid $i$ and its partial pressure in the gas phase, respectively. $P_0$ is a reference pressure which we choose in this work to be the atmospheric pressure without loss of generality. $p(\phi_{vap})$ is a smooth interpolation function commonly used in phase field simulations which ensures the transition from the liquid phase to the vapor phase:

$$p(\phi_{vap}) = \phi_{vap}^2 (3 - 2\phi_{vap}) \tag{6}$$

The last local term of the free energy functional is a purely numerical contribution to ensure numerical stability that helps maintaining the volume fractions between 0 and 1 even for highly incompatible systems or polymer systems, where the equilibrium composition of separated phases might be very close to these boundary values:

$$\Delta G_V^{num}(\{\varphi_i\}) = \sum_{i=1}^n \frac{\beta}{\varphi_i^\gamma} \tag{7}$$

$\beta$ and $\gamma$ are numerical coefficients. $\beta$ is chosen as small as possible, in order to guarantee numerical stability but without modifying significantly the physical behavior of the simulations. The non-local contribution of the free energy represents the contribution of surface tension, which originates from volume fraction gradients and from phase changes:

$$\Delta G_V^{nonloc}\left(\{\nabla\varphi_i\},\{\nabla\phi_{vap}\}\right) = \sum_{i=1}^{n}\frac{\kappa_i}{2}(\nabla\varphi_i)^2 + \frac{\varepsilon_{vap}^2}{2}\left(\nabla\phi_{vap}\right)^2 \tag{8}$$

$\kappa_i$ is the surface tension parameter for the concentration gradient of material $i$ and $\varepsilon_{vap}$ is the surface tension parameters for the gradient of the order parameter representing the liquid-vapor (LV) transition. Note that especially in the case of polymer mixtures, $\kappa_i$ is supposed to include an entropic part that is composition dependent, and that is related to the chain properties and to the interaction parameters [43]. However, we use the $\kappa_i$ as free, constant parameters in this work for simplicity. This is a very common simplification in phase field modelling because its impact on the simulation of liquid-liquid phase separation is practically negligible [44].

## Kinetic equations

Classically, the chemical potential $\mu_i$ of a fluid $i$ in the mixture can be defined as [45]

$$\mu_i = G_m^{loc} + \frac{\partial G_m^{loc}}{\partial x_i} - \sum_{j=1}^{n} x_j \frac{\partial G_m^{loc}}{\partial x_j} \tag{9}$$

In this equation, $G_m^{loc}$ is the local molar Gibb's free energy and $\{x_i\}$ the mole fraction of the fluid $i$. We can extend this definition to incorporate a non-local surface tension term, using functional derivatives instead of partial derivatives. Transforming the equation in terms of volume fractions and free energy densities, and applying this to the free energy density of mixing $\Delta G_V = \Delta G_V^{loc} + \Delta G_V^{nonloc}$, a generalized chemical potential $\mu_i^{gen}$ and a generalized chemical potential density $\mu_{V,i}^{gen}$ can be defined as

$$\mu_{V,i}^{gen} = \frac{\mu_i^{gen}}{N_i v_0} = \left(\Delta G_V + \frac{\delta\Delta G_V}{\delta\varphi_i} - \sum_{j=1}^{n}\varphi_j\frac{\delta\Delta G_V}{\delta\varphi_j}\right) \tag{10}$$

If the molar volume is supposed to be constant, the thermodynamic driving force for the evolution of the volume fractions is the exchange chemical potential density which reads, for all fluids from $1$ to $n-1$ as

$$\mu_{V,j}^{gen} - \mu_{V,n}^{gen} = \frac{\delta\Delta G_V}{\delta\varphi_j} - \frac{\delta\Delta G_V}{\delta\varphi_n} = \frac{\partial\Delta G_V}{\partial\varphi_j} - \frac{\partial\Delta G_V}{\partial\varphi_n} - \left(\nabla\left(\frac{\partial\Delta G_V}{\partial(\nabla\varphi_j)}\right) - \nabla\left(\frac{\partial\Delta G_V}{\partial(\nabla\varphi_n)}\right)\right) \tag{11}$$

The kinetic equation describing the evolution is the so-called Cahn-Hilliard equation, proposed by Cahn and Hilliard for binary mixtures [46][47] and generalized later for multicomponent mixtures [35][48][49]

$$\frac{\partial\varphi_i}{\partial t} = \frac{v_0}{RT}\nabla\left[\sum_{j=1}^{n-1}\Lambda_{ij}\nabla\left(\mu_{V,j}^{gen} - \mu_{V,n}^{gen}\right)\right] \qquad i = 1\dots n-1 \tag{12}$$

This equation is the general version of the diffusion equation for a multicomponent mixture, with the Onsager mobility coefficients being symmetric, $\Lambda_{ij} = \Lambda_{ji}$. It can be shown that the classical diffusion equation is recovered from the Cahn-Hilliard equation when some simplifying assumptions are used (see Supporting Information, S1).

Here, we make a distinction between the mobility coefficients in the liquid phase $\Lambda_{ij}^{liq}$, and the ones in the gas phase $\Lambda_{ij}^{vap}$, while the total mobility is interpolated between both as

$$\Lambda_{ij} = \left(1 - \phi_{vap}\right)\Lambda_{ij}^{liq} + \phi_{vap}\Lambda_{ij}^{vap} \tag{13}$$

In the liquid phase, the mobility has to depend not only on the diffusion coefficients but also on the local mixture composition in order to ensure the incompressibility constraint and the Gibbs-Duhem relationship. Several theories have been proposed to derive correct expressions for the flux, among which the "slow mode theory" [43] and the "fast-mode theory" [50] are the most successful ones. Their names come from the fact that the mutual diffusion coefficient in a binary system is controlled by the slowest component in the "slow-mode theory", while it is controlled by the fastest component in the "fast-mode theory". Since the controversy between both theories is not fully resolved yet despite of significant efforts [51][52][53][54][55], both have been implemented in the model. The expressions of the mobility coefficients in the liquid phase read, for the fast mode

$$\begin{cases} \Lambda_{ii}^{liq} = (1-\varphi_i)^2\omega_i + \varphi_i{}^2 \sum_{k=1,k\neq i}^{n} \omega_k \\ \Lambda_{ij}^{liq} = -(1-\varphi_i)\varphi_j\omega_i - (1-\varphi_j)\varphi_i\omega_j + \varphi_i\varphi_j \sum_{k=1,k\neq i\neq j}^{n} \omega_k \end{cases} \qquad (14)$$

and for the slow mode

$$\begin{cases} \Lambda_{ii} = \omega_i\left(1 - \dfrac{\omega_i}{\sum_{k=1}^{n}\omega_k}\right) \\ \Lambda_{ij} = -\dfrac{\omega_i\omega_j}{\sum_{k=1}^{n}\omega_k} \end{cases} \qquad (15)$$

Here, the coefficients $\omega_i$ are related to the self-diffusion coefficients $D_{s,i}^{liq}$ of the materials $i$ through the relationship $\omega_i = N_i\varphi_i D_{s,i}^{liq}$. The self-diffusion coefficients themselves are usually dependent on the mixture composition, $D_{s,i}^{liq} = D_{s,i}^{liq}(\{\varphi_i\})$. However, unless otherwise specified, they are kept constant in this paper for simplicity, and their values chosen so high that diffusion is fast enough to prevent the formation of any concentration gradients. In such conditions, the choice of the fast or slow mode theory has no impact on the results.

In the gas phase, the composition dependence of the mutual diffusion coefficients and the coupling between fluxes are known to be weak so that we assume the mobility coefficients to be

$$\begin{cases} \Lambda_{ii}^{vap} = \varphi_i D_i^{vap} \\ \Lambda_{ij}^{vap} = 0 \end{cases} \qquad (16)$$

Here, $D_i^{vap}$ is the Fickian diffusion coefficient of the gas in the air. The multiplication by $\varphi_i$ in the equation above compensates the thermodynamic factor $\partial^2\Delta G_V^{vap}/\partial\varphi_i{}^2$ in order to recover the classical Fickian diffusion.

The evolution of the order parameter is given by the classical Allen-Cahn equation,

$$\frac{\partial\phi_{vap}}{\partial t} = -\frac{v_0}{RT}M_{vap}\frac{\delta\Delta G_V}{\delta\phi_{vap}} = -\frac{v_0}{RT}M_{vap}\left(\frac{\partial\Delta G_V}{\partial\phi_{vap}} - \nabla\left(\frac{\partial\Delta G_V}{\partial(\nabla\phi_{vap})}\right)\right) \qquad (17)$$

Here, $M_{vap}$ is the mobility coefficient for the LV interface and will be called "interfacial mobility" in the following. It is thus not to be confused with the LV interface displacement speed in the simulation, as will be detailed below.

The Cahn-Hilliard and the Allen-Cahn equations together ensure that the system progressively relaxes towards its thermodynamic equilibrium by minimizing its free energy relative to the volume fractions and the order parameter. A major simplification in this approach is that the density and molar volume of each material is constant and homogeneous over the whole simulation box, equal to the value in the liquid phase. This requires four comments. First, the reason for this simplification is that it allows for a limited complexity of the equations: for instance, for the calculation of the driving force $\nabla\left(\mu_{V,j}^{gen} - \mu_{V,n}^{gen}\right)$ in the Cahn-Hilliard equations, the use of equations (9) and (10) leads to the very simple equation (11) because some terms simplify. This equation is not valid anymore in the case of inhomogeneous molar volumes. Second, this simplification is a good first order approximation in the liquid phase although not exactly correct. Note that the impact of molar volume or density variations can be partly taken into account through the expression of the diffusion coefficient, for instance using the free volume theory[56]. Third, the density and molar volumes in the gas phase of the simulation are orders of magnitude too high, so that we do not pretend simulating correctly the mass transport in the gas phase using the phase field equations. The impact of the mass transport in the gas phase on the evaporation kinetics is rather taken into account through the expression of the evaporation flux at the boundary of the simulation box. Fourth, this simplification implies that the substantial variations from the liquid phase to the gas phase are not taken into account. Below, we propose a simple correction to the evaporation procedure in order to compensate for this simplification, and demonstrate that the expected drying kinetics can be recovered despite of this extreme simplification.

## *Evaporation procedure and material types*

So far, we have considered the evolution of a closed system. To simulate evaporation, we introduce an outflux boundary condition at each node at the top of the simulation box $z = z_{max}$ where the fluids are in the vapor state (**Figure 1**). The expression of the quantity of solvent $i$ leaving the box $j_i^{z=z_{max}}$ will be discussed in detail in the next section. We define three classes of materials that can be present and which differ through the parameters used to represent them:

- The solvents are defined by their vapor pressure available from experimental data. For simulations of solvent evaporation, they leave the simulation box thanks to the boundary condition $j_i^{z=z_{max}}$.

- The solutes are the non-volatile species. The quantity of solute in the system should be constant, thus the outflux is set to zero for them, $j_i^{z=z_{max}} = 0$. Moreover, in such a phase-field framework, the volume fractions of solutes in the gas phase cannot be zero. To make sure that the overall proportion of solutes being present in the gas phase is negligible as compared to the one in the liquid film, solutes are defined as volatile species, but with a very low vapor pressure.

- The air: for the simulation of solvent evaporation, since the Cahn-Hilliard is a conservation equation, the solvent leaving the simulation box has to be replaced by an additional material that we define as the air. No flux is defined at the boundary for the air since the volume fraction of air is obtained from the conservation of volume, $\sum \varphi_i = 1$. The air is supposed to stay in the gas phase, but a residual quantity of air has to be present in the liquid film. To minimize this effect, and although air at room temperature is far beyond the critical point, we define the air as a material with a very high vapor pressure compared to all other materials. The Flory-Huggins interaction parameters of the air with all other materials in the liquid phase are then set to zero, but the air contributes to the ideal free energy of mixing according to equation (4).

Finally, and most importantly, the simulation is expected to match the Hertz-Knudsen theory. It has already been emphasized that evaporation is a two-step process. First, a phase transformation occurs generating a gas layer in quasi-static equilibrium with the liquid film. This is represented in the simulation by the order parameter transition from 0 to 1 determined by the Allen-Cahn equation. Second, gas molecules diffuse away from this layer, which is represented by the outflux in the simulation. The fact that the evaporation rate is only related to the diffusion step suggests that it is the limiting one, in other words that the phase transition step is very fast compared to diffusion away from the Knudsen layer. This is accounted for in the simulation by choosing a very high interfacial mobility $M_{vap}$ in the Allen-Cahn equation (17), so that the LV interface reacts very quickly to any composition change due to the outflux and restores the quasi-static equilibrium.

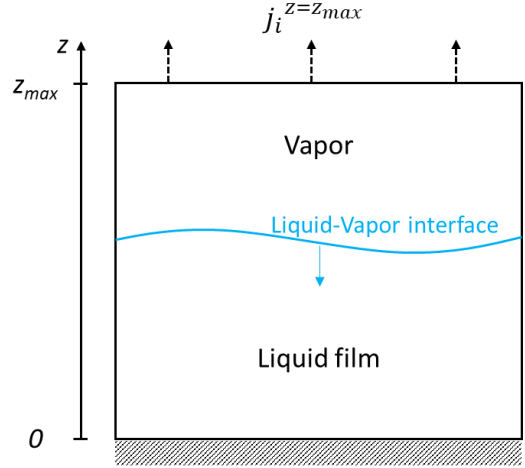

***Figure 1.*** *Setup of the simulation box with the substrate at the bottom ($z = 0$), the solvent outflux at the top boundary ($z = z_{max}$), and the LV interface in-between, separating the drying film from the vapor phase*

## *Expression of the evaporation flux*

Rewriting the Hertz-Knudsen relationship for each solvent $i$ as a volume flux using $M_i = \rho_i N_i v_0$, and assuming that the evaporation and condensation coefficients are equal and have the same value for all solvents, $\alpha_e = \alpha_c = \alpha$, we get the following expression:

$$j_{i,HK} = \alpha \sqrt{\frac{v_0}{2\pi RT}} \frac{N_i}{\rho_i} P_0 \left( \varphi_i^{vap} - \varphi_i^{\infty} \right) \tag{18}$$

In this equation, $\varphi_i^{\infty}$ is defined as $\varphi_i^{\infty} = P_i^{\infty}/P_0$ with $P_i^{\infty}$ being the partial pressure in the environment. $\varphi_i^{vap}$ is the mean volume fraction of the fluid $i$ in the gas phase. According to the Hertz-Knudsen picture, the volume fraction in the gas phase just on top of the LV interface should be used, but for realistic parameters the diffusion in the gas phase is very fast and it turns out that the composition in the vapor is homogeneous, so that $\varphi_i^{vap}$ can be evaluated by simply taking the value at the top boundary.

However, due to the strongly simplifying assumptions of constant density and constant molar volume embedded in the Cahn-Hilliard equation, two adjustments have to be made in order to use the expression given by equation (18). The first adjustment is related to the fact that the quasi-static equilibrium calculated by the model at each time step is not the correct one. On the one hand, the expected equilibrium volume fractions in the liquid and vapor phases can be determined from the free energy density: the equilibrium is defined by equality of the chemical potentials in both phases $\mu_i^{liq} = \mu_i^{vap}$ for each fluid. Using equation (9) and using $\sum \varphi_i = 1$ in both phases, we obtain for instance for a binary mixture the known equation [57]

$$\varphi_1^{vap} = \varphi_{sat,1} \varphi_1^{liq} e^{\left(\left(1-\frac{N_1}{N_2}\right)\left(1-\varphi_1^{liq}\right)+N_1\chi\left(1-\varphi_1^{liq}\right)^2\right)} \qquad (19)$$

whereby $\varphi_1^{vap}$ is the equilibrium volume fraction of the fluid $1$ in the vapor phase and $\varphi_1^{liq}$ in the liquid phase. The expression for $\varphi_2^{vap}$ is the same with permutation of indices $1$ and $2$. On the other hand, the equilibrium calculated by the model results from the condition $\mu_{V,i}^{liq} = \mu_{V,i}^{vap}$. Using equation (10), this results in equilibrium volume fractions calculated in the simulation, $\varphi_{i,simu}^{vap}$, that are related to the expected volume fractions $\varphi_i^{vap}$ by

$$\varphi_i^{vap} = \varphi_{sat,i}\left(\frac{\varphi_{i,simu}^{vap}}{\varphi_{sat,i}}\right)^{N_i} \qquad (20)$$

Below we illustrate this behavior for a binary system. The theoretically expected results (equation (19) full lines) are compared to the values obtained from simulations of binary, closed systems. The simulations are initialized with a homogeneous volume fraction and a LV interface separating liquid and vapor. We let the system equilibrate and measure the equilibrium volume fraction of both fluids in the gas phase $\varphi_{i,simu}^{vap}$. The corrected equilibrium volume fractions $\varphi_i^{vap}$ are then calculated through equation (20) (symbols). **Figure 2** shows the result of this procedure for various values of the molar volumes $N_i$ and of the interaction parameter between both solvents.

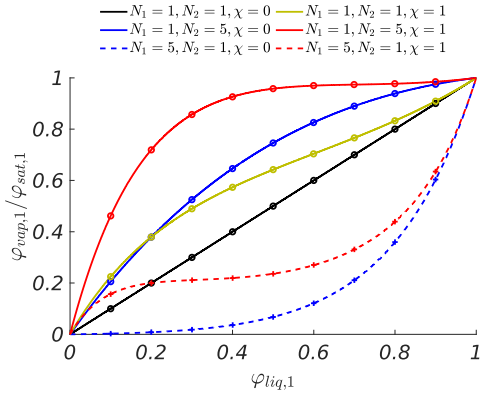

***Figure 2.*** *Equilibrium volume fraction in the gas phase versus the volume fraction in the liquid phase for the fluid 1; comparison of theoretical values (full lines) and values deduced from simulations (symbols).*

For perfectly miscible solvents of the same size, the liquid mixture is ideal and we recover the well-known Raoult behavior. However, as soon as the solvents are of different molar sizes, or are not perfectly miscible, the solution is not ideal anymore and the solubility curve deviates from Raoult's law, generating Henry-like behavior which is more pronounced with increasing molar size difference and/or interaction parameter. This is a direct consequence of the use of the Flory-Huggins theory for the free energy of mixing in the liquid. For solvents with the same molar volume, the solubility curves are identical for both solvents, while they are distinct as soon as the molar volumes become different. Note that for $N_2 = 5$ and $\chi = 1$, the volume fraction of the fluid $1$ (the smallest component) is already almost constant over a liquid volume fraction ranging from 0.5 to 1. The simulation results are perfectly consistent with the theoretical predictions, which demonstrates that the expected LV equilibrium can be derived from the phase-field simulation although the density and molar volume variations are ignored. As a consequence, inserting equation (20) in equation (18), we adjust the expression of the outflux as

$$j_{i,HK} = \alpha\sqrt{\frac{v_0}{2\pi RT}\frac{N_i}{\rho_i}}P_0\left(\varphi_{sat,i}\left(\frac{\varphi_{i,simu}^{vap}}{\varphi_{sat,i}}\right)^{N_i} - \varphi_i^\infty\right) \qquad (21)$$

The second adjustment is due to the fact that in our model the densities are simplified to be constant even upon phase change. As a consequence, the volume of solvent contained in the gas phase is not always negligible compared to the one in the liquid film, as illustrated in **Figure 3**, which significantly changes the mass balance for the solvent. This has to be compensated for in order to recover the Hertz-Knudsen behavior. Let us define the outflux for the solvent $i$ at each node of the upper boundary of the simulation box as $j_i^{z=z_{max}}$. Integrating this outflux at the upper boundary, the total volume variation $dV_i$ of the solvent $i$ in the box is

$$\frac{dV_i}{dt} = dxdy\sum_{z=z_{max}}j_i^{z=z_{max}} \qquad (22)$$

If the volume fraction of solvents in the gas phase would be zero, this mass loss would be entirely converted to a mean displacement of the LV interface with the velocity $v_{int}$. Using $j_i^{z=z_{max}} = j_{i,HK}$, we would get

$$v_{int} = \sum_{i \in \{solv\}} \sum_{z=z_{max}} j_{i,HK} = \alpha \sum_{i \in \{solv\}} \sum_{z=z_{max}} \sqrt{\frac{v_0}{2\pi RT} \frac{N_i}{\rho_i}} P_0 \left( \varphi_{sat,i} \left( \frac{\varphi_{i,simu}^{vap}}{\varphi_{sat,i}} \right)^{N_i} - \varphi_l^{\infty} \right) \tag{23}$$

which is the desired Hertz-Knudsen evaporation rate for a fluid mixture.

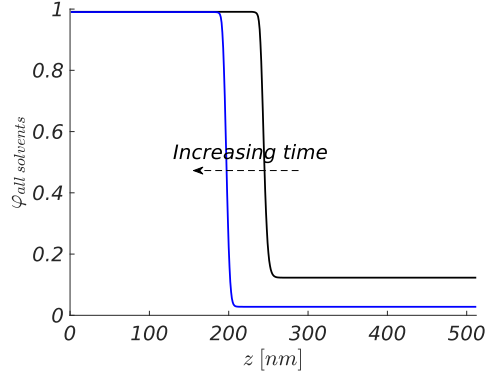

*Figure 3*. Snapshot of the total solvent volume fraction at two different times $t_1$ *(black) and $t_2$ (blue) during evaporation. The liquid phase is on the left and the gas phase on the right and the L-V interface is moving to the left.*

However, with non-negligible quantities of solvent in the gas phase (see **Figure 3**), the outflux has to generate not only the desired mass loss of the liquid film $Sj_{i,HK}$ (S being the surface of the simulation box, $S = \sum_z dx dy$), but also the mass increase $Sv_{int}\varphi_{i,simu}^{vap}$ of solvent in the gas phase due to the displacement of the interface to the left, and the mass loss of the solvent $S(v_{int} + z_{max}\Gamma_{vap}/dt)\Delta\varphi_{i,simu}^{vap}$ in the gas phase due to its volume fraction variation. Here, $\Delta\varphi_{i,simu}^{vap}$ is the volume fraction variation in the vapor phase during one time step and $\Gamma_{vap}$ is the proportion of vapor phase in the whole box and thus $z_{max}\Gamma_{vap}$ the mean vapor height in the box. This mass balance leads to the final expression of the outflux implemented at the upper boundary of the simulation box:

$$j_l^{z=z_{max}} = j_{i,HK} - (\varphi_{i,simu}^{vap} + \Delta\varphi_{i,simu}^{vap}) \left( \sum_{k \in \{solv\}} j_{k,HK} \right) + \frac{z_{max}\Gamma_{vap}}{dt} \Delta\varphi_{i,simu}^{vap} \tag{24}$$

## *Computational Details*

The coupled kinetic equations (12) to (17) with the boundary conditions (equations (21) and (24)), and using the free energy density defined by equations (2) to (8) are implemented with a finite volume scheme. They are written in a dimensionless form using $\widetilde{\Delta G_V^{loc}} = \Delta G_V^{loc}/g_{sc}$, $\widetilde{\Delta G_V^{nonloc}} = \Delta G_V^{nonloc}/(g_{sc}l_{sc}^2)$, $\tilde{l} = l/l_{sc}$, $\widetilde{\Lambda}_{ij} = \Lambda_{ij}/D_{sc}$ and $\tilde{t} = t/t_{sc}$. The coefficients $g_{sc}$, $l_{sc}$, $D_{sc}$ and $t_{sc}$ are chosen as $g_{sc} = RT/v_0$, $l_{sc} = \sqrt{max(\kappa_{1...n}, \varepsilon_{vap}^2)} / g_{sc}$ to be consistent with the size of the thinnest interface of the system, $D_{sc} = max(N_i D_{s,i})$ and finally $t_{sc} = l_{sc}^2/D_{sc}$.

The equations are numerically solved simultaneously using an Euler backward implicit scheme with variable time steps, which is the main numerical improvement compared to our previous work[26]. For this, both Allen-Cahn and Cahn-Hilliard equations are linearized and solved together with a direct solver, the Cahn-Hilliard equation being written in the split form[58]. This implicit implementation is a crucial pre-requisite to perform such simulations with a complete decoupling of several different time scales, namely the very fast diffusion processes in the simulated domain (related to the Cahn-Hilliard equation), the still fast build-up of the LV equilibrium at the film surface (related to the Allen-Cahn equation), and the very slow evaporation process (related to the outflux). Indeed, an explicit implementation would suffer from drastic limitations of the time steps in order to ensure numerical stability. If using realistic input parameters, this would imply unaffordable computational time, even for small 1D-simulations like the ones presented in this work and despite any parallel implementation. Inversely, the Euler backward method is unconditionally stable and allows for the use of much larger time steps.

We use a simple heuristic strategy to adapt the time step: it is of course required that all volume fractions everywhere in the simulation box lie in the $]0,1[$ interval (0 and 1 being mathematically excluded, see for instance equation (4)). If this condition is fulfilled with the calculated solution, the time step is increased by 20% for the next time increment. Otherwise, the time increment is rejected and recalculated with a twice smaller time step. Moreover, an upper limit is set to the time step with respect to the Courant-Friederichs-Lewy (CFL) criterion calculated using the expected interface velocity (equation (23)), $\Delta t < C_{CFL}\Delta x/v_{int}(t)$.

Our code is natively three-dimensional and implemented in parallel, but the simulations shown in this work are one-dimensional simulations with a lattice of 512 nodes (pure solvent evaporation, two solvent evaporation) or 2048 lattice nodes (solute deposition) performed on a single CPU core. The lattice resolution is $\Delta x = 1 nm$ and the CFL criterion calculated with $0.03 < C_{CFL} < 0.3$. Under these conditions, the simulations require a few minutes for 500 to 15000 time steps. It has been verified that they were converged in time by checking that the results remain unchanged with a significantly more restrictive CFL-criterion. The spatial convergence has also been verified on the test cases with the thinnest LV interfaces by using various lattice resolutions. The parameters used in the simulation, unless specified in text, are summarized in Table 1. $P_{sat,air}$ and $P_{sat,solute}$ are handled as adjustable parameters that are chosen respectively as high and as low as possible, so that the volume fraction of air in the film and of solute in the gas phase be as low as possible, while ensuring a stable numerical resolution. Likewise, we choose $\varepsilon_{vap}$ and the $\kappa_i$ parameters as small as possible in order to obtain a numerically tractable interface thickness with a grid spacing of 1nm in all simulations presented here (unless specified otherwise). The LV interface profile can be quite sharp with volume fractions coming close to 0 and 1, and the numerical contribution to the free energy (equation (7)) helps stabilizing the resolution, even with the low value of the coefficient $\beta$ chosen here.

| $T$ | 300 K |
|---|---|
| $\rho_i$ (all) | 1000 kg/m$^3$ |
| $v_0$ | $3 \cdot 10^{-5}$ m$^3$/mol |
| $N_{air}$ | 1 |
| $N_i$ (others) | See text |
| $\chi$ | See text |
| $P_{sat,air}$ | $10^8$ Pa |
| $P_{sat,solute}$ | $10^2$ Pa |
| $P_{sat}$ (others) | See text |
| $P_i^\infty$ | 0 Pa |
| $\kappa_i$ (all) | $6 \cdot 10^{-10}$ J/m |
| $\varepsilon_{vap}$ | $9 \cdot 10^{-5}$ (J/m)$^{0.5}$ |
| $D_{s,i}^{liq}$ (all) | $2 \cdot 10^{-9}$ m$^2$/s |
| $D_i^{vap}$ (all) | $10^{-5}$ m$^2$/s |
| $M_{vap}$ | $10^6$ s$^{-1}$ |
| $\alpha$ | $-2.3 \cdot 10^{-5}$ |
| $\beta$ | $10^{-7}$ |
| $\gamma$ | 1 |

Table 1. Basic parameter set for the simulations

# Results

## *Evaporation of pure solvents*

We first turn to the simulation of the evaporation of a single solvent. The system is composed of the solvent itself plus the air. The typical volume fraction field and order parameter field in such a situation are shown in **Figure 4**. As expected, the solvent volume fraction in the gas phase is homogeneous, constant and equal to $P_{sat}/P_0$. We checked that, as expected, the solvent evaporates with constant evaporation rate until it disappears almost completely from the simulation domain (not shown). This shows that the LV quasi-static equilibrium is reached at any time during the evaporation. Additionally, it has been verified that once the interfacial mobility $M_{vap}$ is sufficiently high, the evaporation rate is fully insensitive to the surface tension parameters $\kappa$, $\varepsilon_{vap}$ and to $M_{vap}$ itself (not shown). This validates our approach and confirms that the quasi-static equilibrium condition is fulfilled and that the evaporation is diffusion limited as depicted in the Hertz-Knudsen theory. This proves also the choice of concentration-independent $\kappa_i$ coefficients as an acceptable simplification here.

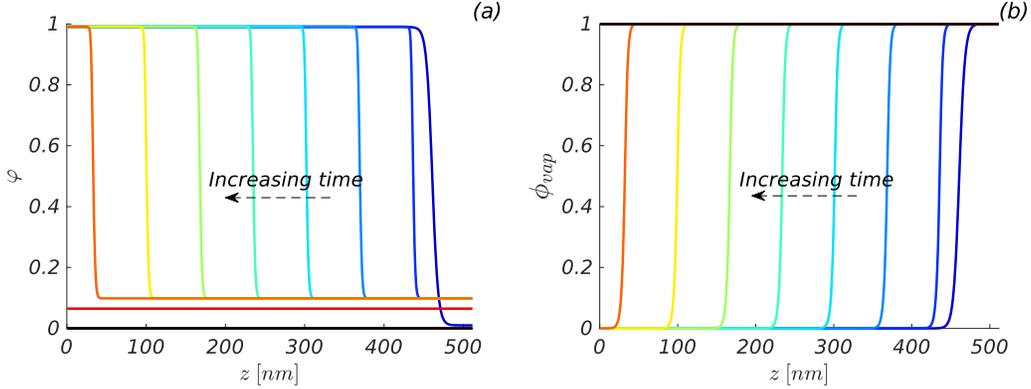

**Figure 4.** *Typical volume fraction field (a) and order parameter field (b) for a single solvent system during evaporation. The interface moves from right to left from the initial state (deep blue) to the final state (black). From the second curve, the time interval between the different curves is constant. From deep blue to orange, the liquid film is present on the left and becomes thinner with increasing drying time. At the red curve, the solvent has fully evaporated and only the gas phase remains with a certain amount of solvent. In the final state, the solvent fully disappears from the simulation box.*

Therefore, we expect the evaporation rate to be exactly given by the Hertz-Knudsen relationship. This is evidenced by the results plotted in **Figure 5**: the product $P_{sat}\sqrt{N}$ has been varied over four orders of magnitude and the simulated evaporation rates almost match the expected values. A deviation is observed for both curves $N = 1$ and $N = 5$ when the vapor pressure and hence the volume fraction in the gas phase is very low. This is due to the numerical contribution to the free energy (equation (7)) which is not negligible anymore and modifies the equilibrium. For pure solvents, this is not a problem since $P_0$ can be chosen so that the volume fraction in the gas phase is in a satisfactory range. For simulations of solvent blends, however, this implies that the vapor pressures of all solvents may differ by roughly three decades at most. We believe that this is not restrictive for practical cases.

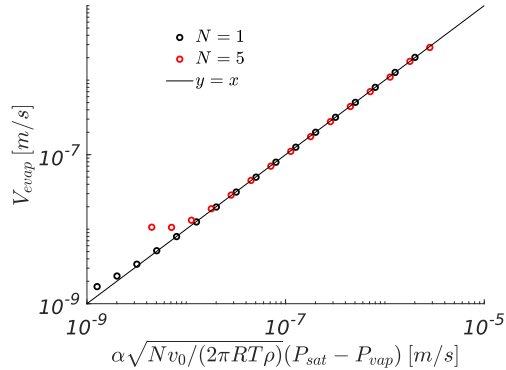

**Figure 5.** *Simulated evaporation rates of pure solvents versus the expected evaporation rates from equation (23), for various vapor pressures and two molar sizes.*

As a consequence, the evaporation rate from our simulations is perfectly known and given by equation (23) and since the coefficient $\alpha$ is not allowed to vary with the solvent, we can compare the results to experimentally measured evaporation rates for different solvents. For this, we use the data from Ref [59]. Thereby, the mass evaporation rates is being converted to volume evaporation rates. The partial pressure in the environment is assumed to be negligible. The results are shown in **Figure 6**, relative to the evaporation rate of toluene. This is in fact a check of the validity of the Hertz-Knudsen relationship, which shows a good agreement with the experimental data with some discrepancies. Our simulations behave identically and possible differences between simulation and experiments might be due to the fact that we use a very simple version of the Hertz-Knudsen relationship. The refinement of this theory reviewed in Ref. [29] can be easily accounted for by simply changing the expression of the outflux (equation (18)) in the future, but this goes beyond the scope of the current paper.

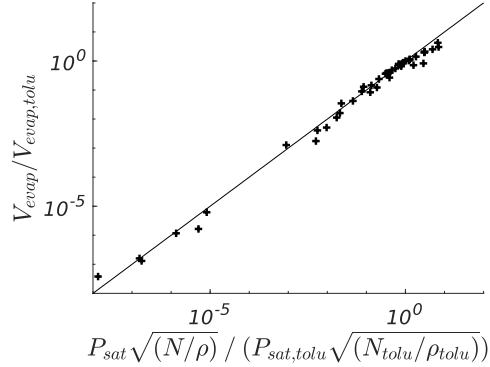

**Figure 6.** *Correlation between measured evaporation rates (from Ref. [59]) of various solvents (relative to toluene) and simulated evaporation rates / theoretical predictions of equation (23).*

Finally, we investigate the surface tension dependence of the solvents on their vapor pressure. This is known experimentally to vary smoothly with $ln(P_{sat})$. We calculate the surface tension $\sigma$ from the simulations at different vapor pressures by the classical van der Waals formula applied to the volume fraction and order parameter gradients (all other parameters stay constant):

$$\sigma = (\kappa_1 + \kappa_2) \int \left(\frac{d\varphi}{dz}\right)^2 dz + \varepsilon_{vap}^2 \int \left(\frac{d\phi_{vap}}{dz}\right)^2 dz \qquad (25)$$

Here, the integral is taken over the interface from the one phase to the other. The comparison with experimental data taken from [59] and for a selection of solvent used in organic photovoltaics (data from the Hansen's solvent database [60]) is shown in **Figure 7**. Quantitatively, the simulated surface tensions are more than one order of magnitude higher than the experimental values. This is expected since the surface tension parameters have been adjusted to generate interface thicknesses of at least 5-6nm that can be resolved with the chosen 1nm grid spacing. This is a well-known drawback of phase-field simulations, and this is absolutely not a problem here, even for quantitative simulations, since the evaporation rates are fully independent of the LV interface properties. Qualitatively, the simulated surface tension follows the experimental tendency very nicely. Nevertheless, a deviation can be seen at high vapor pressure, which can be explained easily: in this range, the simulated solvent volume fractions in the vapor phase are far from zero, leading to an underestimation of the volume fraction gradients and hence of the surface tension. The simulated surface tension values presented in **Figure 7** have been calculated with the same values of $\kappa_i$ for all solvents, whereas it is expected to be solvent-dependent. This explains the considerable deviations between simulated and experimental values. In principle, this deviations can be suppressed by mapping the $\kappa_i$ parameters of the solvents to the experimental surface tensions.

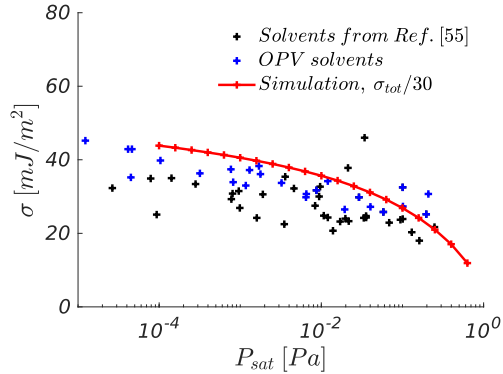

**Figure 7.** *Comparison of the simulated surface tension dependence on the vapor pressure ($N_{solvent} = 5$) with experimental data for a selection of solvents.*

## Evaporation of fluid mixtures: solvent blends

To investigate the evaporation behavior of fluid mixtures, we first simulate the evaporation of a binary solvent blend without solute. For the first solvent, the molar volume is set equal to three times the one of the air ($N = 3$), and the vapor pressure is fixed to the one of toluene at room temperature (20°C, 2800 Pa, $\varphi_{sat} = 0.028$). The liquid film is a 50:50 blend at the beginning of the simulation. We vary the vapor pressure, the molar volume of the second solvent, as well as the interaction parameter between both solvents. Compared to the case of a pure solvent, the evaporation rate is expected to vary with time as soon as the solvents are different or not perfectly miscible.

We first confirm again that the simulation produces a ternary solvent-solvent-air LV quasi-static equilibrium at any time. To do this, we solve the ternary equilibrium expected from the condition $\mu_{V,i}^{liq} = \mu_{V,i}^{vap}$ numerically (see Supporting information, S2) and compare to the volume fractions given by the simulation at each time step. We find a perfect agreement (not shown).

We then compare the simulated evaporation rates to the ones that are expected for a *binary* (without air) solvent-solvent equilibrium. To obtain the theoretical curve, we use the kinetic equation of the interface displacement which is given by

$$\frac{dh}{dt} = \sum_{i=1}^{2} v_i(t) = \sum_{i=1}^{2} \alpha \sqrt{\frac{v_0}{2\pi RT} \frac{N_i}{\rho_i}} P_0(\varphi_{vap,i}(\varphi_i^{liq}) - \varphi_i^{\infty}) \tag{26}$$

where $h$ is the film height and $\varphi_{vap,i}(\varphi_i^{liq})$ is given by equation (19). In parallel, the evolution of the total volume of both fluids $V_1$ and $V_2$ is given by

$$\frac{dV_i}{dt} = Sv_i \tag{27}$$

with $S$ being the film surface. The volume fraction variation of the first fluid in the liquid phase is given by

$$\varphi_1^{liq} = \frac{V_1}{V_1 + V_2} \tag{28}$$

To compare with the simulation results, we numerically integrate equations (26-28) from the initial conditions used in the simulation. The results of this procedure are shown in **Figure 8** for various vapor pressures and the molar volumes of the second solvent, as well as different interaction parameters between both solvents.

Except when both solvents are identical and perfectly compatible (black curve), the overall evaporation rate is in general time dependent. This effect finds its origin in the constantly changing LV equilibrium: with two solvents evaporating at different speeds, the composition of the liquid film is constantly changing. For solvents of the same molar volume (blue curve), the overall evaporation rate, which is determined at the beginning by the fastest evaporating solvent (the one with the highest vapor pressure), slowly decreases to reach the evaporation rate of the slowest solvent when the fastest has disappeared. For solvents with different molar volumes and identical vapor pressures, the "fastest" evaporating solvent, if it were alone, is the one with the highest molar volume due to the $\sqrt{N_i}$ prefactor in the Hertz-Knudsen formula. However, in the blend, due to the LV equilibrium, the partial pressure of this solvent is substantially lower, as illustrated in **Figure 2**. As a result, the evaporation rate of this "fastest" evaporating solvent is lower, its volume fraction increases progressively and therefore is the overall evaporation rate (green curve). When both molar volume and vapor pressure of the solvents are distinct, the effect of the vapor pressure can compensate the effect of the molar volume and the evaporation rate can be either increasing or decreasing.

The simulation results are in very good agreement with the theoretical expectations. This proves again that the Hertz-Knudsen picture of solvent evaporation is reproduced, and that the presence of the air in the simulation does not significantly perturb the calculated quasi-static LV equilibrium. For the simulations with $P_{sat2} = 5P_{sat1}$, we recover the theoretical results despite a very significant amount of solvent of the gas phase (total volume fraction in the range 10-15% at the beginning of the simulation). This shows that the corrections to the outflux deriving from the mass balance in the box work perfectly. We checked that the model behavior is still correct with volume fractions in the gas phase up to 50% or higher (not shown). We also check again that the drying kinetics is independent of the LV interface profile, and hence of the parameter $\kappa_i$ and $\varepsilon_{vap}$, as well as of $M_{vap}$ provided it is high enough (not shown).

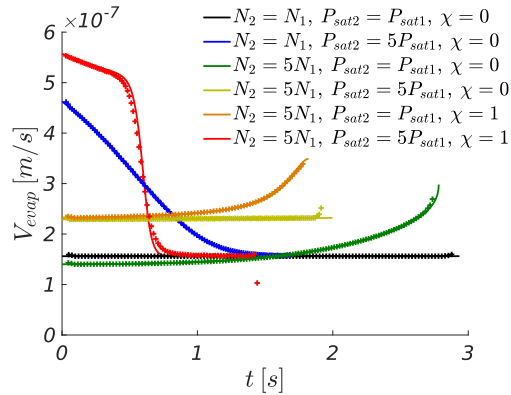

**Figure 8.** *Evaporation kinetics of a binary solvent blend for various parameters for the second solvent and the interaction parameter. For the 1$^{st}$ solvent, $N = 3$ and $\varphi_{sat} = 0.028$. Theory, equations (26-28) (full lines) and simulation (symbols).*

The evaporation rate is time-dependent because it is composition-dependent, and the film composition varies with time. Therefore, the evolution of the film composition should be correct, which we illustrate in **Figure 9**. Beyond the drying kinetics, this might be crucial for the proper simulation of the film structuration: for example, critical processes such as liquid-liquid phase separation or nucleation are only triggered from a given composition in the film. Once again, the agreement between theory and simulation is excellent.

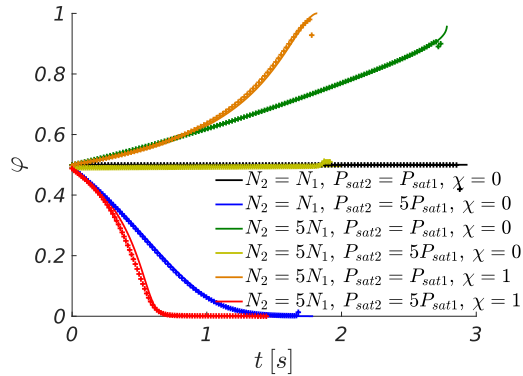

***Figure 9.*** *Volume fraction of the second solvent in the liquid film during evaporation of a binary blend, for various parameters for the second solvent and the interaction parameter. For the $1^{st}$ solvent, $N = 3$ and $\varphi_{sat} = 0.028$. Theory, equations (26-28) (full lines) and simulation (symbols).*

In practical situations, solvent blending is known as a powerful tool in order to monitor the structuration of the drying film. The blend is often chosen depending on criteria such as the interactions with the various solutes, their wetting properties and their individual evaporation rate. However, for a solvent blend, the evaporation kinetics and the composition of the drying film has a non-trivial time-evolution, depending on the molar volumes, densities and vapor pressures of the individual solvents. This in turn can also have an influence on the film structuring and should be taken into account for an optimal choice of the solvent blend. We hope that this kind of model can help for such considerations.

## *Evaporation of fluid mixtures: solute deposition*

Towards deposition of solution-processed thin films, we finally investigate the case of a drying mixture containing one solvent and one solute. The three fluids in the simulation are the solute, the solvent and the air, as described earlier. We only study the impact of the solution thermodynamics on the time dependence of the evaporation rate. In order to do so, we perform simulations at low Biot numbers. The Biot number $Bi = \frac{h(t)v_{int}(\varphi(t))}{D_m(\varphi(t))}$, where $D_m$ is the mutual diffusion coefficient, is in general concentration- and time-dependent, but we chose sufficiently high diffusion coefficients in the liquid phase to ensure $Bi \ll 1$ during the whole drying. This ensures that diffusion processes are sufficiently fast to compensate for concentration gradients that could be generated by the displacement of the LV interface. Otherwise, the gas layer would be in equilibrium with the upper layer of the liquid film whose composition is not equal to the mean composition. This is known to have a strong impact on the time-dependence of the evaporation rate [26 35] and is deliberately avoided here. However, this purely kinetic effect will be illustrated later in this paper on the case of a PS-toluene mixture.

The typical field evolution in such a simulation is shown in **Figure 10**. The solvent volume fraction in the vapor phase is once again homogeneous and decreases with time. The volume fraction field in the liquid film is also homogeneous, except at the film surface. The brown curve shows the final, dry state which does not change further with time. The solvent volume fraction peak is not due to the kinetic effect described above, the Biot number in our simulation being below $10^{-4}$, but this is simply the equilibrium profile of the LV interface. It is due to the fact that the solvents have a much higher vapor pressure than the solutes, and thus preferentially occupy the interface. The height of the peak is at first order determined by the ratio of the vapor pressures of the solvent and the solute. We are not sure that this peak is physical but it could make sense that solvent molecules preferentially gather at the very surface before/while undergoing the LV phase transition. Luckily, as already stated several times above, the interface profile has no impact on the kinetics, except the following small bias: the interface (and hence the peak) is unrealistically broad in order allow for numerically tractable simulations. As a result, coming closer and closer to the dry state, the amount of solvent contained in this region might become important compared to the amount in the bulk liquid film. This might result in slightly modified equilibrium compositions and differences between simulation and theory. Simply increasing the box size can solve this problem, so that the volume at the interface becomes negligible compared to the volume in the bulk. For this reason, we performed solute deposition simulations with 2048 grid points, leading to a final film height of 180nm as shown in **Figure 10**. The fields for the case with 512 grid points (final height of 45nm) are shown in the Supporting information (S3) for comparison.

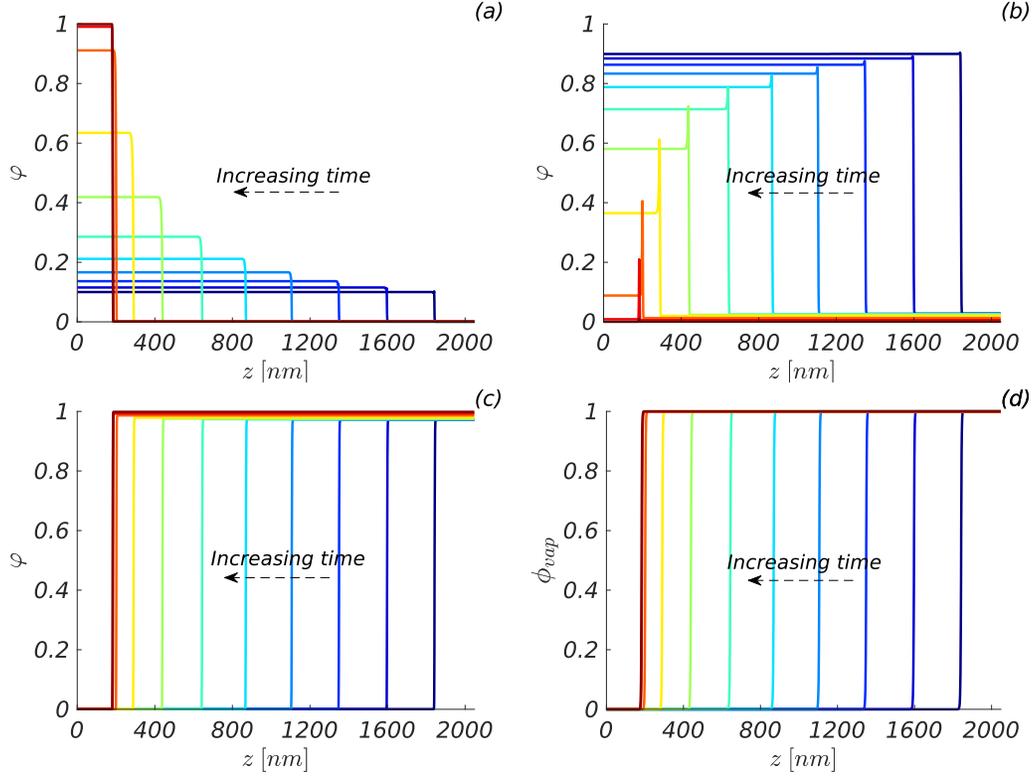

***Figure 10.*** *Typical volume fraction field for the solute (a), the solvent (b), the air (c) and order parameter field (d) during drying. The interface moves from right to left from the initial state (blue) to the final state (brown). The system is discretized by 2048 points.*

Here again, we analyze the time-dependence of the evaporation rate for various parameters and compare it to the theoretical results obtained with the method detailed in the previous paragraph (**Figure 11**). The time-dependence of the liquid film composition, which is the reason for the changes in the drying rate, has also been analyzed (**Figure 12**). Once again, the agreement between theory and experiment is very good. For the ideal mixture (black curve), the evaporation rate decreases very smoothly and the equation of the curve can be derived analytically [26 35 61]. The more incompatible the solute and the solvent are and the bigger the solute molecules are, the more constant is the evaporation rate. This can be qualitatively understood going back to **Figure 2**. For high interaction parameters and different molecular sizes, the solubility curve of the smaller component is shifted so that the equilibrium partial pressure of the solvent becomes almost constant for a broad range of volume fractions in the liquid film and therefore, for a long period of the drying process in the simulation.

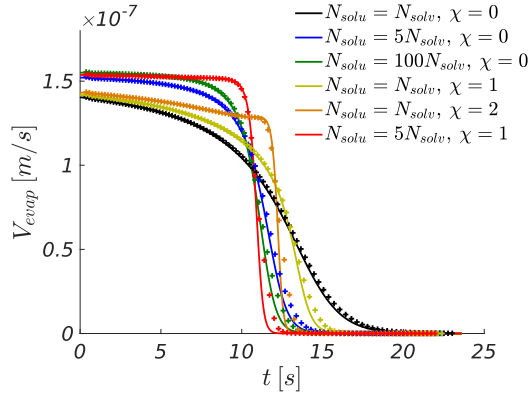

***Figure 11.*** *Evaporation kinetics of a binary solute-solvent mixture for various model parameters from the theory (full lines) and simulation (symbols). For the solvent, $N = 3$ and $\varphi_{sat} = 0.028$.*

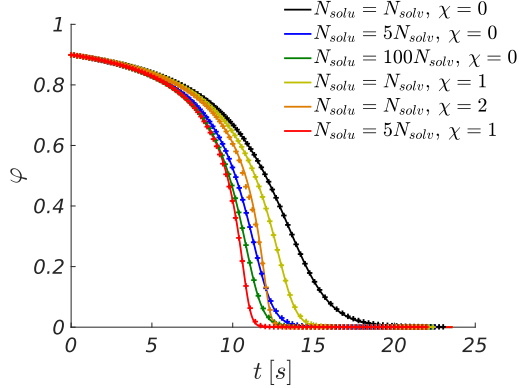

***Figure 12.*** *Volume fraction of the solvent during evaporation of a binary solute-solvent mixture for various model parameters. Theory (full lines) and simulation (symbols). For the solvent, $N = 3$ and $\varphi_{sat} = 0.028$.*

This is expected to be a general phenomenon, and leads to the conclusion that for a given solvent, this thermodynamic effect pushes the evaporation rate of a polymer solution to be much more constant than the one of a small molecule solution. This produces the so-called "constant rate" drying phase of polymer solutions. However, the low diffusion coefficients at high polymer concentrations at the end of the drying might lead to a high Biot number and strong concentration gradients in the film. The kinetic effect becomes dominant as compared to the thermodynamic effect and the drying rate can drop over several order of magnitudes [36]. Similarly, poorly compatible solutions should tend to evaporate at a more constant rate, at least as long as they stay miscible and that no liquid-liquid phase separation is triggered.

## Comparison with experimental results on the case of polystyrene-toluene

For a quantitative illustration, we simulate the experiments of drying polystyrene-toluene solutions reported in our previous work[26]. The parameters for toluene are taken from the Hansen's solvent database[60]. The experiments were performed at room temperature (20°C) and films of about 2 micrometers dry thickness were produced. The molecular weight of the linear polystyrene was chosen deliberately low at 35kg/mol to avoid the build-up of concentration gradients as long as possible. The self-diffusion coefficients of polystyrene and toluene in the mixture are known to be strongly composition-dependent. Various data are available from the literature for the self-diffusion coefficients of toluene [34] [38] and of polystyrene [30] [31] [33] at different concentrations, as well as for the mutual diffusion coefficient [30] [31] [34] [62]. Nevertheless, we are not aware of all data being available for this molecular weight at room temperature, and we could only estimate the diffusion properties from the available data and approximated them with simple mathematical functions for the self-diffusion coefficients. The Onsager coefficient and mutual diffusion coefficient is calculated as a result of the slow mode theory (see Supporting information S4). The Flory-Huggins interaction parameter between polystyrene and toluene is estimated to be between *0.3* [30] and *0.45* [63]. This leaves $\alpha$ as the only adjustable parameter influencing the evaporation rate in the model, and it is fitted to the evaporation rate of *0.155μm/s* measured in our experiment, leading to the value $\alpha=2.3 \cdot 10^{-5}$ used throughout the paper. The grid spacing has been chosen to be 10 nanometers; we adjusted $\kappa_i = 6 \cdot 10^{-8}$ (J/m) and $\epsilon_{vap} = 9 \cdot 10^{-4}$ (J/m)$^{0.5}$ in order to have a sufficiently broad LV interface. The agreement between experimental and simulation results, shown in **Figure 13**, is very good and demonstrates the quantitative accuracy of the phase-field simulation framework presented in this paper. The Biot number is of the order of 0.01 at the beginning of the drying and the thermodynamic effect is dominant; the evaporation rate is almost constant, due to the high molecular weight of the polymer compared to the solvent, and drops suddenly at the end of the drying as expected from the LV equilibrium. At the same time, when the volume fraction of solvent has reached roughly 50%, the Biot number gets close to 1 and concentration gradients appear in the film (see Supporting information S4). This leads to a polymer skin at the top of the film with very slow diffusion, so that, when the remaining solvent volume fraction reaches about 10%, the drying rate decreases by more than one order of magnitude. Note that we cannot observe this second phase of slow drying in the experiments, which end shortly after the end of the "constant rate" phase. Unfortunately, we do not have any means of evaluating the residual solvent quantity at the end of the measurements, but the films might not be fully dry. Since we expect this huge slow-down when the volume fraction of solvent reaches 10%, we assume the residual solvent quantity in the experiments to be 10% for the sake of the comparison with the simulation. The behavior at longer times (until the complete drying in the simulation) is shown in the Supporting information S4. Since the measurements are performed below the glass transition temperature of the blend at low solvent concentration (roughly below 10-15%) [38], we expect an impact of the relaxation kinetics that can further slow down the drying kinetics. However, we cannot observe such an effect, probably because the measurement time is not sufficient.

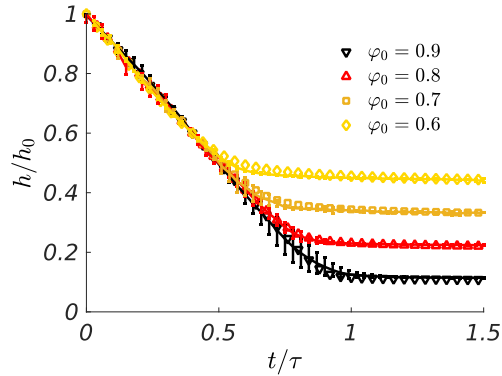

*Figure 13. Simulated time dependent film height (symbols) for different volume fractions $\varphi_0$ of solvent in the initial film, and compared to the experimental results (full lines, reproduced from Ref. [26] with permission from the PCCP Owner Societies) for a 35kg/mol polystyrene – toluene mixture. Error bars represent +/- 1 standard deviation. $h_0$ is the initial height and $\tau$ the time needed to dry a film with $\varphi_0 = 0.9$ at constant drying rate.*

## *Example of 2D simulations with surface deformation*

One of the key features of our framework is that the LV interface is located inside the simulation box. This allows the tracking of its position without any further changes to the model, even if the interface deforms or bends. In particular, no remeshing technique is required and a simple fixed, rectangular, regular mesh can be used. To illustrate this feature, we perform 2D simulations of a drying solvent-solute binary blend on a rough substrate (see **Figure 14**). We use a regular 256*128 mesh with a grid spacing of 2nm (the detailed parameters of the simulation are given in the Supporting information, S5). Two structures of different heights are present on the substrate. Several simulations with an initial solute volume fraction varying from 0 to 0.8 are performed. For the highest solute volume fractions ($\varphi_0 = 0.8$ and $\varphi_0 = 0.6$), the dry film fully covers both structures. For $\varphi_0 = 0.4$, it can be seen that the interface bends around the highest structure so that the dry film is not completely flat. For $\varphi_0 = 0.2$, the interface bends around both substrate structures. Finally, for $\varphi_0 = 0$ (pure solvent drying), the substrate is fully dewetted and the final interface follows the substrate morphology. Note that in that case, dewetting in the flat area between both structures occurs before the end of the drying, so that the LV interface separates into two distinct parts (not shown). This demonstrates the potential of our framework to handle different situations where surface deformation has to be taken into account. Not only drying on structured, rough substrates can be considered but also any kind of situation where the morphology formation during drying might lead to interface deformations or even to a rough film (for instance during drying of immiscible polymer mixtures [64] or crystallizing films). This will be the topic of intensive future work.

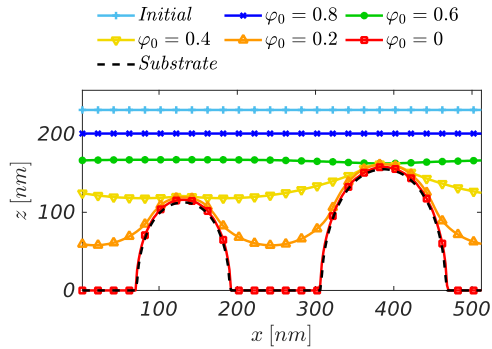

*Figure 14: initial (light blue) and final position of the film surface after drying of a solute-solvent mixture on a rough substrate. The final state of different simulations with initial solute volume fractions varying from $\varphi_0 = 0.8$ (dark blue) to $\varphi_0 = 0$ (red) are shown. The substrate surface is indicated by the dashed black line.*

## Conclusion and perspectives

In this work, we developed a phase-field simulation framework that mimics the Hertz-Knudsen description of evaporation processes. The fast interfacial mobility used in the Allen-Cahn equation ensures a quasi-static equilibrium between the liquid phase and the vapor phase close to the film surface. The diffusion of gas molecules away from the film surface is modeled with a flux boundary condition that recovers the Hertz-Knudsen relationship for the evaporation rate of a pure solvent. The solvent surface tension is also successfully reproduced. For fluid mixtures, the simple situation where the drying is not influenced by gas phase resistances, limited diffusion in the liquid film, or polymer relaxation, has been investigated. Our evaporation procedure generates drying curves in excellent agreement with the theoretical results that can be expected from the LV-equilibrium calculated with the Flory-Huggins theory. The simulations have also successfully been compared to experimental data obtained in such a simple case and we obtained a nice alignment of simulation, theory and measurements. The simulation framework can handle surface deformation and hence film roughness without any further modification, as has been exemplified in 2D simulations of drying on a rough substrate.

In this paper, in order to test the validity of the phase-field model, we showed simulations of the very simple situation where the LV equilibrium together with the Hertz-Knudsen approach is responsible for the drying kinetics. However, in real systems, as detailed before, other physical processes may play an important role and the model can be improved in several ways to take them into account:

- Since we used a simple Hertz-Knudsen theory, for pure solvent evaporation our simulations inherit from all its advantages, but also from its imperfections documented by comparisons with experimental results. We believe that this could be improved in a straightforward manner by implementing the progress made on the Hertz-Knudsen theory [29] in the formula for the flux at the boundary.

- Similarly, the role of the processes in the gas phase (Stefan flow, forced laminar or even turbulent convection) can be integrated by simply making the coefficient $\alpha$, or more generally the expression of the outflux (equation (18)), dependent on the nature and intensity of the gas flow using classical phenomenological descriptions like Sherwood correlations [65].

- For the evaporation kinetics of solvent and polymer mixtures, potential deviations from experiments could be associated with the accuracy of the Flory-Huggins theory and corrected easily by an improved description of the thermodynamics of the liquid solution (typically composition-dependent interaction parameters, other forms for the excess energy of mixing such as the Redlich-Kistler extension [45]).

- For film drying simulations, our model predicts solvent accumulation in the LV interface. Whether this accumulation is physical or just a consequence of the diffuse interface approach, as well as the consequences on the simulation results, are questions that should be addressed in near future.

In the future, such an improved model could be cross-checked with exhaustive already available data. However, it has already been shown that these situations could be successfully calculated in the one-dimensional case [32] [36] [37]. In fact, the benefits of such a tool become clear in much more complex situations. Possible examples of these complex situations are mixtures of 4, 5 or more fluids which is after all quite common in the solution-processing of thin films: since our framework is readily written for any number of components in the mixtures, they could be in principle simulated without any further modifications of the governing equations. Other examples are films where evaporation-induced phase transformation (spinodal decomposition, crystallization…) takes place and leads to strong inhomogeneity and time-dependence of the film properties. An example of a practical application is solution-processed organic photoactive layers, for which both aspects of complexity are relevant. The numerically efficient implementation of the code allows for heavy 2D/3D simulations of such situations. This will be demonstrated in future work, coupling the equations presented here with previous work [66].

To the best of our knowledge, the existing simulation procedures used to study the structure formation upon drying in two or three dimensions most often describe only the liquid film and set an outflux at the upper surface of the film. This outflux is sometimes assumed to be constant [67] or proportional to the liquid volume fraction at the surface [16]. In this context, we believe that the best solution is the one proposed by Dehsari and co-workers where the solvent partial pressures are calculated as $P_{sat}e^{\mu/RT}$ [68], which produces in fact the same evaporation flux as compared to our approach. This should lead to more accurately simulated drying kinetics, drying times, and therefore final morphologies, and beyond this to a better understanding of the dry film properties. Nevertheless, we believe that our model brings one interesting feature: even using a simple, regular fixed mesh, the film surface deformation can be tracked and effects such as roughness formation or incomplete substrate coverage can be accounted for, similar to what has been already demonstrated [26] [64]. This will be of highest importance for the simulation of film drying on rough substrates, as well as for situations where the morphology formation during evaporation results in rough dry films. This can typically be the case when liquid-liquid phase separation occurs in polymer systems, or when crystallization processes occur during drying. For example, this is a well-known problem for the quality of the photoactive layer in perovskite solar cells [7] [69] [70].

# Author information

**Corresponding Author**

E-mail: o.ronsin@fz-juelich.de (O.R.)
E-mail: j.harting@fz-juelich.de (J.H.)

**ORCID**

Olivier Ronsin: 0000-0002-3958-8636
Jens Harting: 0000-0002-9200-6623
Hans-Joachim Egelhaaf: 0000-0002-8263-8125
Christoph Brabec: 0000-0002-9440-0253

**Notes**

The authors declare no competing financial interest.

# Acknowledgments

The authors acknowledge financial support by the German Research Foundation (DFG, projects HA 4382/14-1 and BR 4031/20-1) and gratefully thank Dr. Olga Wodo for fruitful discussions.

# Associated content

Supporting Information: derivation of the Fick equation and of the ternary LV equilibrium, additional details to the simulations of solute deposition and to the 2D simulations.

# Literature

(1) Eslamian, M. Inorganic and Organic Solution-Processed Thin Film Devices. *Nano-Micro Letters* **2017**, *9* (1), 3. https://doi.org/10.1007/s40820-016-0106-4.

(2) Brabec, C. J.; Durrant, J. R. Solution-Processed Organic Solar Cells. *MRS Bulletin* **2008**, *33* (7), 670–675. https://doi.org/10.1557/mrs2008.138.

(3) Williams, S. T.; Rajagopal, A.; Chueh, C.-C.; Jen, A. K.-Y. Current Challenges and Prospective Research for Upscaling Hybrid Perovskite Photovoltaics. *The Journal of Physical Chemistry Letters* **2016**, *7* (5), 811–819. https://doi.org/10.1021/acs.jpclett.5b02651.

(4) Li, J.; Daniel, C.; Wood, D. Materials Processing for Lithium-Ion Batteries. *Journal of Power Sources* **2011**, *196* (5), 2452–2460. https://doi.org/10.1016/j.jpowsour.2010.11.001.

(5) Mehta, V.; Cooper, J. S. Review and Analysis of PEM Fuel Cell Design and Manufacturing. *Journal of Power Sources* **2003**, *114* (1), 32–53. https://doi.org/10.1016/S0378-7753(02)00542-6.

(6) Hellmann, C.; Treat, N. D.; Scaccabarozzi, A. D.; Razzell Hollis, J.; Fleischli, F. D.; Bannock, J. H.; de Mello, J.; Michels, J. J.; Kim, J.-S.; Stingelin, N. Solution Processing of Polymer Semiconductor: Insulator Blends-Tailored Optical Properties through Liquid-Liquid Phase Separation Control. *J. Polym. Sci. Part B: Polym. Phys.* **2015**, *53* (4), 304–310. https://doi.org/10.1002/polb.23656.

(7) Moore, D. T.; Sai, H.; Tan, K. W.; Smilgies, D.-M.; Zhang, W.; Snaith, H. J.; Wiesner, U.; Estroff, L. A. Crystallization Kinetics of Organic–Inorganic Trihalide Perovskites and the Role of the Lead Anion in Crystal Growth. *Journal of the American Chemical Society* **2015**, *137* (6), 2350–2358. https://doi.org/10.1021/ja512117e.

(8) Wadsworth, A.; Hamid, Z.; Kosco, J.; Gasparini, N.; McCulloch, I. The Bulk Heterojunction in Organic Photovoltaic, Photodetector, and Photocatalytic Applications. *Advanced Materials* **2020**, *32*, 2001763. https://doi.org/10.1002/adma.202001763.

(9) Ye, L.; Collins, B. A.; Jiao, X.; Zhao, J.; Yan, H.; Ade, H. Miscibility-Function Relations in Organic Solar Cells: Significance of Optimal Miscibility in Relation to Percolation. *Advanced Energy Materials* **2018**, *8* (28), 1703058. https://doi.org/10.1002/aenm.201703058.

(10) Cui, Y.; Yao, H.; Zhang, J.; Zhang, T.; Wang, Y.; Hong, L.; Xian, K.; Xu, B.; Zhang, S.; Peng, J.; Wei, Z.; Gao, F.; Hou, J. Over 16% Efficiency Organic Photovoltaic Cells Enabled by a Chlorinated Acceptor with Increased Open-Circuit Voltages. *Nat Commun* **2019**, *10* (1), 2515. https://doi.org/10.1038/s41467-019-10351-5.

(11) Cui, Y.; Yao, H.; Zhang, J.; Xian, K.; Zhang, T.; Hong, L.; Wang, Y.; Xu, Y.; Ma, K.; An, C.; He, C.; Wei, Z.; Gao, F.; Hou, J. Single-Junction Organic Photovoltaic Cells with Approaching 18% Efficiency. *Advanced Materials* **2020**, *32* (19), 1908205. https://doi.org/10.1002/adma.201908205.

(12) Machui, F.; Maisch, P.; Burgués-Ceballos, I.; Langner, S.; Krantz, J.; Ameri, T.; Brabec, C. J. Classification of Additives for Organic Photovoltaic Devices. *ChemPhysChem* **2015**, *16* (6), 1275–1280. https://doi.org/10.1002/cphc.201402734.

(13) Yang, M.; Li, Z.; Reese, M. O.; Reid, O. G.; Kim, D. H.; Siol, S.; Klein, T. R.; Yan, Y.; Berry, J. J.; van Hest, M. F. A. M.; Zhu, K. Perovskite Ink with Wide Processing Window for Scalable High-Efficiency Solar Cells. *Nature Energy* **2017**, *2* (5), 17038. https://doi.org/10.1038/nenergy.2017.38.

(14) Lee, C.-K.; Pao, C.-W. Multiscale Molecular Simulation of Solution Processing of SMDPPEH: PCBM Small-Molecule Organic Solar Cells. *ACS Applied Materials & Interfaces* **2016**, *8* (32), 20691–20700. https://doi.org/10.1021/acsami.6b05027.

(15) Kipp, D.; Ganesan, V. Influence of Block Copolymer Compatibilizers on the Morphologies of Semiflexible Polymer/Solvent Blends. *The Journal of Physical Chemistry B* **2014**, *118* (16), 4425–4441. https://doi.org/10.1021/jp501207y.

(16) Wodo, O.; Ganapathysubramanian, B. Modeling Morphology Evolution during Solvent-Based Fabrication of Organic Solar Cells. *Computational Materials Science* **2012**, *55*, 113–126. https://doi.org/10.1016/j.commatsci.2011.12.012.

(17)    Negi, V.; Wodo, O.; van Franeker, J. J.; Janssen, R. A. J.; Bobbert, P. A. Simulating Phase Separation during Spin Coating of a Polymer–Fullerene Blend: A Joint Computational and Experimental Investigation. *ACS Appl. Energy Mater.* **2018**, *1* (2), 725–735. https://doi.org/10.1021/acsaem.7b00189.

(18)    Schaefer, C.; Michels, J. J.; van der Schoot, P. Structuring of Thin-Film Polymer Mixtures upon Solvent Evaporation. *Macromolecules* **2016**, *49* (18), 6858–6870. https://doi.org/10.1021/acs.macromol.6b00537.

(19)    Saylor, D. M.; Forrey, C.; Kim, C.-S.; Warren, J. A. Diffuse Interface Methods for Modeling Drug-Eluting Stent Coatings. *Ann Biomed Eng* **2016**, *44* (2), 548–559. https://doi.org/10.1007/s10439-015-1375-7.

(20)    Michels, J. J.; Zhang, K.; Wucher, P.; Beaujuge, P. M.; Pisula, W.; Marszalek, T. Predictive Modelling of Structure Formation in Semiconductor Films Produced by Meniscus-Guided Coating. *Nature Materials* **2021**, *20* (1), 68–75. https://doi.org/10.1038/s41563-020-0760-2.

(21)    Cirillo, E. N. M.; Colangeli, M.; Moons, E.; Muntean, A.; Muntean, S.-A.; van Stam, J. A Lattice Model Approach to the Morphology Formation from Ternary Mixtures during the Evaporation of One Component. *Eur. Phys. J. Spec. Top.* **2019**, *228* (1), 55–68. https://doi.org/10.1140/epjst/e2019-800140-1.

(22)    Hessling, D.; Xie, Q.; Harting, J. Diffusion Dominated Evaporation in Multicomponent Lattice Boltzmann Simulations. *The Journal of Chemical Physics* **2017**, *146* (5), 054111. https://doi.org/10.1063/1.4975024.

(23)    Badillo, A. Quantitative Phase-Field Modeling for Boiling Phenomena. *Physical Review E* **2012**, *86* (4), 041603. https://doi.org/10.1103/PhysRevE.86.041603.

(24)    Kaempfer, T. U.; Plapp, M. Phase-Field Modeling of Dry Snow Metamorphism. *Physical Review E* **2009**, *79* (3), 031502. https://doi.org/10.1103/PhysRevE.79.031502.

(25)    Schweigler, K. M.; Ben Said, M.; Seifritz, S.; Selzer, M.; Nestler, B. Experimental and Numerical Investigation of Drop Evaporation Depending on the Shape of the Liquid/Gas Interface. *International Journal of Heat and Mass Transfer* **2017**, *105*, 655–663. https://doi.org/10.1016/j.ijheatmasstransfer.2016.10.033.

(26)    Ronsin, O. J. J.; Jang, D.; Egelhaaf, H.-J.; Brabec, C. J.; Harting, J. A Phase-Field Model for the Evaporation of Thin Film Mixtures. *Phys. Chem. Chem. Phys.* **2020**, *22* (12), 6638–6652. https://doi.org/10.1039/D0CP00214C.

(27)    Hertz, H. Ueber Die Verdunstung Der Flüssigkeiten, Insbesondere Des Quecksilbers, Im Luftleeren Raume. *Annalen der Physik* **1882**, *253* (10), 177–193. https://doi.org/10.1002/andp.18822531002.

(28)    Knudsen, M. Die Maximale Verdampfungsgeschwindigkeit Des Quecksilbers. *Annalen der Physik* **1915**, *352* (13), 697–708. https://doi.org/10.1002/andp.19153521306.

(29)    Persad, A. H.; Ward, C. A. Expressions for the Evaporation and Condensation Coefficients in the Hertz-Knudsen Relation. *Chemical Reviews* **2016**, *116* (14), 7727–7767. https://doi.org/10.1021/acs.chemrev.5b00511.

(30)    Krüger, K.-M.; Sadowski, G. Fickian and Non-Fickian Sorption Kinetics of Toluene in Glassy Polystyrene. *Macromolecules* **2005**, *38* (20), 8408–8417. https://doi.org/10.1021/ma050353o.

(31)    Zettl, U.; Hoffmann, S. T.; Koberling, F.; Krausch, G.; Enderlein, J.; Harnau, L.; Ballauff, M. Self-Diffusion and Cooperative Diffusion in Semidilute Polymer Solutions As Measured by Fluorescence Correlation Spectroscopy. *Macromolecules* **2009**, *42* (24), 9537–9547. https://doi.org/10.1021/ma901404g.

(32)    Schabel, W.; Scharfer, P.; Kind, M.; Mamaliga, I. Sorption and Diffusion Measurements in Ternary Polymer–Solvent–Solvent Systems by Means of a Magnetic Suspension Balance—Experimental Methods and Correlations with a Modified Flory–Huggins and Free-Volume Theory. *Chemical Engineering Science* **2007**, *62* (8), 2254–2266. https://doi.org/10.1016/j.ces.2006.12.062.

(33)    Liu, R.; Gao, X.; Adams, J.; Oppermann, W. A Fluorescence Correlation Spectroscopy Study on the Self-Diffusion of Polystyrene Chains in Dilute and Semidilute Solution. *Macromolecules* **2005**, *38* (21), 8845–8849. https://doi.org/10.1021/ma0511090.

(34)    Zielinski, J. M.; Duda, J. L. Predicting Polymer/Solvent Diffusion Coefficients Using Free-Volume Theory. *AIChE J.* **1992**, *38* (3), 405–415. https://doi.org/10.1002/aic.690380309.

(35)    Schaefer, C. *Theory of Nanostructuring in Solvent-Deposited Thin Polymer Films*; Technische Universiteit Eindhoven, 2016, 9-36.

(36)    Siebel, D.; Scharfer, P.; Schabel, W. Prediction of Diffusion in a Ternary Solvent–Solvent–Polymer Blend by Means of Binary Diffusion Data: Comparison of Capillary and Simulative Results. *Journal of Applied Polymer Science* **2016**, *133* (36), 43899. https://doi.org/10.1002/app.43899.

(37)    Merklein, L.; Eser, J. C.; Börnhorst, T.; Könnecke, N.; Scharfer, P.; Schabel, W. Different Dominating Mass Transport Mechanisms for Drying and Sorption of Toluene-PMMA Films – Visualized with Raman Spectroscopy. *Polymer* **2021**, *222*, 123640. https://doi.org/10.1016/j.polymer.2021.123640.

(38)    Mueller, F.; Naeem, S.; Sadowski, G. Toluene Sorption in Poly(Styrene) and Poly(Vinyl Acetate) near the Glass Transition. *Ind. Eng. Chem. Res.* **2013**, *52* (26), 8917–8927. https://doi.org/10.1021/ie302322t.

(39)    Börnhorst, T.; Scharfer, P.; Schabel, W. Drying Kinetics from Micrometer- to Nanometer-Scale Polymer Films: A Study on Solvent Diffusion, Polymer Relaxation, and Substrate Interaction Effects. *Langmuir* **2021**, *37* (19), 6022–6031. https://doi.org/10.1021/acs.langmuir.1c00641.

(40)    Schmidt-Hansberg, B.; Klein, M. F. G.; Peters, K.; Buss, F.; Pfeifer, J.; Walheim, S.; Colsmann, A.; Lemmer, U.; Scharfer, P.; Schabel, W. *In Situ* Monitoring the Drying Kinetics of Knife Coated Polymer-Fullerene Films for Organic Solar Cells. *Journal of Applied Physics* **2009**, *106* (12), 124501. https://doi.org/10.1063/1.3270402.

(41)    Gu, Z.; Alexandridis, P. Drying of Films Formed by Ordered Poly(Ethylene Oxide)−Poly(Propylene Oxide) Block Copolymer Gels. *Langmuir* **2005**, *21* (5), 1806–1817. https://doi.org/10.1021/la0495130.

(42)    Flory, P. J. *Principles of Polymer Chemistry*; Cornell University Press, 1953, 495-539.

(43)  de Gennes, P. G. Dynamics of Fluctuations and Spinodal Decomposition in Polymer Blends. *The Journal of Chemical Physics* **1980**, *72* (9), 4756–4763. https://doi.org/10.1063/1.439809.

(44)  Nauman, E. B.; He, D. Q. Nonlinear Diffusion and Phase Separation. *Chemical Engineering Science* **2001**, *56*, 1999–2018.

(45)  Hillert, M. *Phase Equilibria, Phase Diagrams and Phase Transformations: Their Thermodynamic Basis*; Cambridge University Press: Cambridge, UK; New York, 2008, 63-79 & 441-459.

(46)  Cahn, J. W.; Hilliard, J. E. Free Energy of a Nonuniform System. I. Interfacial Free Energy. *The Journal of Chemical Physics* **1958**, *28* (2), 258–267. https://doi.org/10.1063/1.1744102.

(47)  Cahn, J. W. On Spinodal Decomposition. *Acta Metallurgica* **1961**, *9* (9), 795–801. https://doi.org/10.1016/0001-6160(61)90182-1.

(48)  Shang, Y.; Fang, L.; Wei, M.; Barry, C.; Mead, J.; Kazmer, D. Verification of Numerical Simulation of the Self-Assembly of Polymer-Polymer-Solvent Ternary Blends on a Heterogeneously Functionalized Substrate. *Polymer* **2011**, *52* (6), 1447–1457. https://doi.org/10.1016/j.polymer.2011.01.038.

(49)  Huang, C.; de la Cruz, M. O.; Swift, B. W. Phase Separation of Ternary Mixtures: Symmetric Polymer Blends. *Macromolecules* **1995**, *28* (24), 7996–8005. https://doi.org/10.1021/ma00128a005.

(50)  Kramer, E. J.; Green, P.; Palmstrøm, C. J. Interdiffusion and Marker Movements in Concentrated Polymer-Polymer Diffusion Couples. *Polymer* **1984**, *25* (4), 473–480. https://doi.org/10.1016/0032-3861(84)90205-2.

(51)  Composto, R. J.; Kramer, E. J.; White, D. M. Fast Macromolecules Control Mutual Diffusion in Polymer Blends. *Nature* **1987**, *328* (6127), 234–236. https://doi.org/10.1038/328234a0.

(52)  Jablonski, E. L.; Gorga, R. E.; Narasimhan, B. Interdiffusion and Phase Behavior at Homopolymer/Random Copolymer Interfaces. *Polymer* **2003**, *44* (3), 729–741. https://doi.org/10.1016/S0032-3861(02)00826-1.

(53)  Murschall, U.; Fischer, E. W.; Herkt-Maetzky, C.; Fytas, G. Investigation of the Mutual Diffusion in Compatible Mixture of Two Homopolymers by Photon Correlation Spectroscopy. *Journal of Polymer Science Part C: Polymer Letters* **1986**, *24* (4), 191–197. https://doi.org/10.1002/pol.1986.140240408.

(54)  Higgins, J. S.; Fruitwala, H. A.; Tomlins, P. E. Experimental Evidence for Slow Theory of Mutual Diffusion Coefficients in Phase Separating Polymer Blends. *British Polymer Journal* **1989**, *21* (3), 247–257. https://doi.org/10.1002/pi.4980210312.

(55)  Akcasu, A. Z. The "Fast" and "Slow" Mode Theories of Interdiffusion in Polymer Mixtures: Resolution of a Controversy. *Macromolecular Theory and Simulations* **1997**, *6* (4), 679–702. https://doi.org/10.1002/mats.1997.040060401.

(56)  Vrentas, J. S.; Duda, J. L. Diffusion in Polymer—Solvent Systems. I. Reexamination of the Free-Volume Theory. *J. Polym. Sci. Polym. Phys. Ed.* **1977**, *15* (3), 403–416. https://doi.org/10.1002/pol.1977.180150302.

(57)  Wolf, B. A. Making Flory–Huggins Practical: Thermodynamics of Polymer-Containing Mixtures. In *Polymer Thermodynamics*; Wolf, B. A., Enders, S., Eds.; Springer Berlin Heidelberg: Berlin, Heidelberg, 2010; Vol. 238, pp 1–66. https://doi.org/10.1007/12_2010_84.

(58)  Elliott, C. M.; French, D. A.; Milner, F. A. A Second Order Splitting Method for the Cahn-Hilliard Equation. *Numerische Mathematik* **1989**, *54* (5), 575–590. https://doi.org/10.1007/BF01396363.

(59)  Mackay, D.; van Wesenbeeck, I. Correlation of Chemical Evaporation Rate with Vapor Pressure. *Environmental Science & Technology* **2014**, *48* (17), 10259–10263. https://doi.org/10.1021/es5029074.

(60)  Abbott, S.; Hansen, C. M.; Yamamoto, H. *Hansen Solubility Parameters in Practice*, 5th Edition.; Hansen-Solubility.com, 2015, 259-269.

(61)  Ternes, S.; Börnhorst, T.; Schwenzer, J. A.; Hossain, I. M.; Abzieher, T.; Mehlmann, W.; Lemmer, U.; Scharfer, P.; Schabel, W.; Richards, B. S.; Paetzold, U. W. Drying Dynamics of Solution-Processed Perovskite Thin-Film Photovoltaics: In Situ Characterization, Modeling, and Process Control. *Advanced Energy Materials* **2019**, *9* (39), 1901581. https://doi.org/10.1002/aenm.201901581.

(62)  Zielinski, J. M. A Friction Factor Analysis of the Coupling between Polymer/Solvent Self- and Mutual-diffusion: Polystyrene/Toluene. *Journal of Polymer Science: Part B: Polymer Physic* **1996**, *34*, 2759–2766. https://doi.org/10.1002/(sici)1099-0488(19961130)34:163.0.co.

(63)  Interaction Parameter http://polymerdatabase.com/polymer%20physics/Chi%20Table2.html (accessed 2021 -01 -13).

(64)  Cummings, J.; Lowengrub, J. S.; Sumpter, B. G.; Wise, S. M.; Kumar, R. Modeling Solvent Evaporation during Thin Film Formation in Phase Separating Polymer Mixtures. *Soft Matter* **2018**, *14* (10), 1833–1846. https://doi.org/10.1039/C7SM02560B.

(65)  Schmidt-Hansberg, B.; Baunach, M.; Krenn, J.; Walheim, S.; Lemmer, U.; Scharfer, P.; Schabel, W. Spatially Resolved Drying Kinetics of Multi-Component Solution Cast Films for Organic Electronics. *Chemical Engineering and Processing: Process Intensification* **2011**, *50* (5), 509–515. https://doi.org/10.1016/j.cep.2010.12.012.

(66)  Ronsin, O. J. J.; Harting, J. Role of the Interplay between Spinodal Decomposition and Crystal Growth in the Morphological Evolution of Crystalline Bulk Heterojunctions. *Energy Technol.* **2020**, *8* (12), 1901468. https://doi.org/10.1002/ente.201901468.

(67)  Kim, C.-S.; Saylor, D. M.; McDermott, M. K.; Patwardhan, D. V.; Warren, J. A. Modeling Solvent Evaporation during the Manufacture of Controlled Drug-Release Coatings and the Impact on Release Kinetics. *Journal of Biomedical Materials Research Part B: Applied Biomaterials* **2009**, *90B* (2), 688–699. https://doi.org/10.1002/jbm.b.31336.

(68)  Dehsari, H. S.; Michels, J. J.; Asadi, K. Processing of Ferroelectric Polymers for Microelectronics: From Morphological Analysis to Functional Devices. *J. Mater. Chem. C* **2017**, *5* (40), 10490–10497. https://doi.org/10.1039/C7TC01495C.

(69)  Hu, Q.; Zhao, L.; Wu, J.; Gao, K.; Luo, D.; Jiang, Y.; Zhang, Z.; Zhu, C.; Schaible, E.; Hexemer, A.; Wang, C.; Liu, Y.; Zhang, W.; Grätzel, M.; Liu, F.; Russell, T. P.; Zhu, R.; Gong, Q. In Situ Dynamic Observations of Perovskite

Crystallisation and Microstructure Evolution Intermediated from [PbI6]4− Cage Nanoparticles. *Nature Communications* **2017**, *8* (1), 15688. https://doi.org/10.1038/ncomms15688.

(70) Eperon, G. E.; Burlakov, V. M.; Docampo, P.; Goriely, A.; Snaith, H. J. Morphological Control for High Performance, Solution-Processed Planar Heterojunction Perovskite Solar Cells. *Advanced Functional Materials* **2014**, *24* (1), 151–157. https://doi.org/10.1002/adfm.201302090.

# Supporting Information
# "Phase-field simulation of liquid-vapor equilibrium and evaporation of fluid mixtures"

## S1. Reduction of the Cahn-Hilliard equation to the classical Fick equations

The Cahn-Hilliard equation set reads

$$\frac{\partial \varphi_i}{\partial t} = \frac{v_0}{RT} \nabla \left[ \sum_{j=1}^{n-1} \Lambda_{ij} \nabla \left( \mu_{V,j}^{gen} - \mu_{V,n}^{gen} \right) \right] \qquad i = 1 \dots n-1 \tag{S29}$$

Reducing to a binary mixture we obtain a single kinetic equation for $\varphi = \varphi_1 = 1 - \varphi_2$

$$\frac{\partial \varphi}{\partial t} = \frac{v_0}{RT} \nabla \left[ \Lambda_{11} \nabla \left( \mu_{V,1}^{gen} - \mu_{V,2}^{gen} \right) \right] \tag{S30}$$

Dropping the surface tension (gradient) terms in equation (11) of the main text, we get

$$\frac{\partial \varphi}{\partial t} = \nabla \left[ \Lambda_{11} \nabla \left( \frac{1 + ln\varphi}{N_1} - \frac{1 + ln(1 - \varphi)}{N_2} + \chi_{12}(1 - 2\varphi) \right) \right] \tag{S31}$$

Remembering that for any function $f$, we have $\nabla f(\varphi) = \frac{\partial f}{\partial \varphi} \nabla \varphi$, this leads to

$$\frac{\partial \varphi}{\partial t} = \nabla \left[ \Lambda_{11} \left( \frac{1}{N_1 \varphi} + \frac{1}{N_2(1 - \varphi)} - 2\chi_{12} \right) \nabla \varphi \right] \tag{S32}$$

Inside the bracket, we can identify the first law of Fick, with a composition-dependent mutual diffusion coefficient:

$$D_m = \Lambda_{11} \left( \frac{1}{N_1 \varphi} + \frac{1}{N_2(1 - \varphi)} - 2\chi_{12} \right) \tag{S33}$$

Now, if we consider the case of two fluids with the same molar volume ($N_1 = N_2 = 1$) and with the same, concentration-independent self-diffusion coefficient $D_{s,1}^{liq}(\varphi) = D_{s,2}^{liq}(\varphi) = D$, both fast-mode theory and slow-mode theory simplify to the same expression of the Onsager coefficient $\Lambda_{11} = D\varphi(1 - \varphi)$. Assuming in addition that the mixture is ideal ($\chi_{12} = 0$), we obtain a constant mutual diffusion coefficient $D_m = D$ and therefore have

$$\frac{\partial \varphi}{\partial t} = D\Delta\varphi \tag{S34}$$

Using the general relationship $\varphi_i = c_i v_i$ between the volume fraction, the molar volume and the concentration $c_i$ for all materials, this transforms to the classical second law of Fick:

$$\frac{\partial c}{\partial t} = D\Delta c \tag{S35}$$

Note that for $N_1 \neq N_2$:

- in the slow-mode theory, $\Lambda_{11} = \frac{N_1 \varphi_1 D_{s,1}^{liq}(\varphi) N_2 \varphi_2 D_{s,2}^{liq}(\varphi)}{N_1 \varphi_1 D_{s,1}^{liq}(\varphi) + N_2 \varphi_2 D_{s,2}^{liq}(\varphi)}$ and we obtain again the second law of Fick with a constant diffusion coefficient
- in the fast-mode theory, $\Lambda_{11} = \varphi_2^2 \varphi_1 N_1 D_{s,1}^{liq}(\varphi) + \varphi_1^2 \varphi_2 N_2 D_{s,2}^{liq}(\varphi)$, and in order to get a concentration-independent mutual diffusion coefficient, the self-diffusion coefficients have to be assumed to be concentration-dependent using the LBV equation (also known as multicomponent Darken equation)[1], more precisely $D_{s,i}^{liq}(\varphi) = \left( \frac{N_i}{N_j} \frac{\varphi_i}{D} + \frac{\varphi_j}{D} \right)^{-1}$.

# S2. Derivation of the LV equilibrium for a ternary mixture

Here, we show how to calculate the LV equilibrium of a ternary mixture. This is more involved than for a binary mixture and can only be solved analytically if the ternary mixture is ideal. For the general case of a non-ideal mixture, it can be solved numerically using the following procedure: we have in fact 7 unknown variables, namely the three volume fractions in the gas phase, the three volume fractions in the liquid phase, and the overall proportion of the liquid phase. To solve for these 7 variables, we have the 7 following equations:

- two conservation equations $\sum \varphi_i = 1$ in the liquid and the vapor phase, respectively.

- two equations expressing the overall repartition of the fluids between the vapor and the liquid phase for given average volume fractions in the whole system $\bar\varphi_1$ and $\bar\varphi_2$ (remember that $\Gamma_{vap}$ is the proportion of vapor phase in the whole box):

$$\left(1 - \Gamma_{vap}\right)\varphi_i^{liq} + \Gamma_{vap}\varphi_i^{vap} = \bar\varphi_i \tag{S36}$$

- three equations coming from the resolution of the LV equilibrium, $\mu_{V,i}^{liq} = \mu_{V,i}^{vap}$ for each fluid. Using again equation (10) of the main text, we find

$$\varphi_1^{vap} = P_{sat,1}\left(\varphi_1^{liq}\right)^{\frac{1}{N_1}} e^{\left(\frac{1-\varphi_1^{liq}}{N_1} - \frac{\varphi_2^{liq}}{N_2} - \frac{\varphi_3^{liq}}{N_3} + \chi_{12}\left(1-\varphi_1^{liq}\right)\varphi_2^{liq} + \chi_{13}\left(1-\varphi_1^{liq}\right)\varphi_3^{liq} - \chi_{23}\varphi_2^{liq}\varphi_3^{liq}\right)} \tag{S37}$$

and similar expressions for $\varphi_2^{vap}$ and $\varphi_3^{vap}$ with permuted indices. Using the conservation equation for the liquid volume fractions in order to drop $\varphi_3^{liq}$, we can write the three equation (S37) as $\varphi_i^{vap} = f_i\left(\varphi_1^{liq}, \varphi_2^{liq}\right)$, and using the conservation equation in the vapor phase in order to drop $\varphi_3^{vap}$, we get on the one hand

$$f_1\left(\varphi_1^{liq}, \varphi_2^{liq}\right) + f_2\left(\varphi_1^{liq}, \varphi_2^{liq}\right) + f_3\left(\varphi_1^{liq}, \varphi_2^{liq}\right) = 1 \tag{S38}$$

On the other hand, inserting $\varphi_i^{vap} = f_i\left(\varphi_1^{liq}, \varphi_2^{liq}\right)$ in both equations (S36) and taking the ratio of both, we get

$$f_1\left(\varphi_1^{liq}, \varphi_2^{liq}\right)\left[\varphi_2^{liq} - \bar\varphi_2\right] + f_2\left(\varphi_1^{liq}, \varphi_2^{liq}\right)\left[\bar\varphi_1 - \varphi_1^{liq}\right] = \bar\varphi_1\varphi_2^{liq} - \bar\varphi_2\varphi_1^{liq} \tag{S39}$$

Both equations (S38) and (S39) can be solved numerically to in order to get $\varphi_2^{liq}$ depending on $\varphi_1^{liq}$, and the intersection of both curves gives the desired solution for $\varphi_1^{liq}$ and $\varphi_2^{liq}$. To compare this theoretical result to the simulations, this procedure is performed for the time-dependent $\bar\varphi_1$ and $\bar\varphi_2$ extracted from the simulation.

Note that the same procedure can be used to solve for the real equilibrium by using $\mu_i^{liq} = \mu_i^{vap}$ instead of $\mu_{V,i}^{liq} = \mu_{V,i}^{vap}$.

## S3. Typical volume fraction fields for solute-solvent mixture, small system

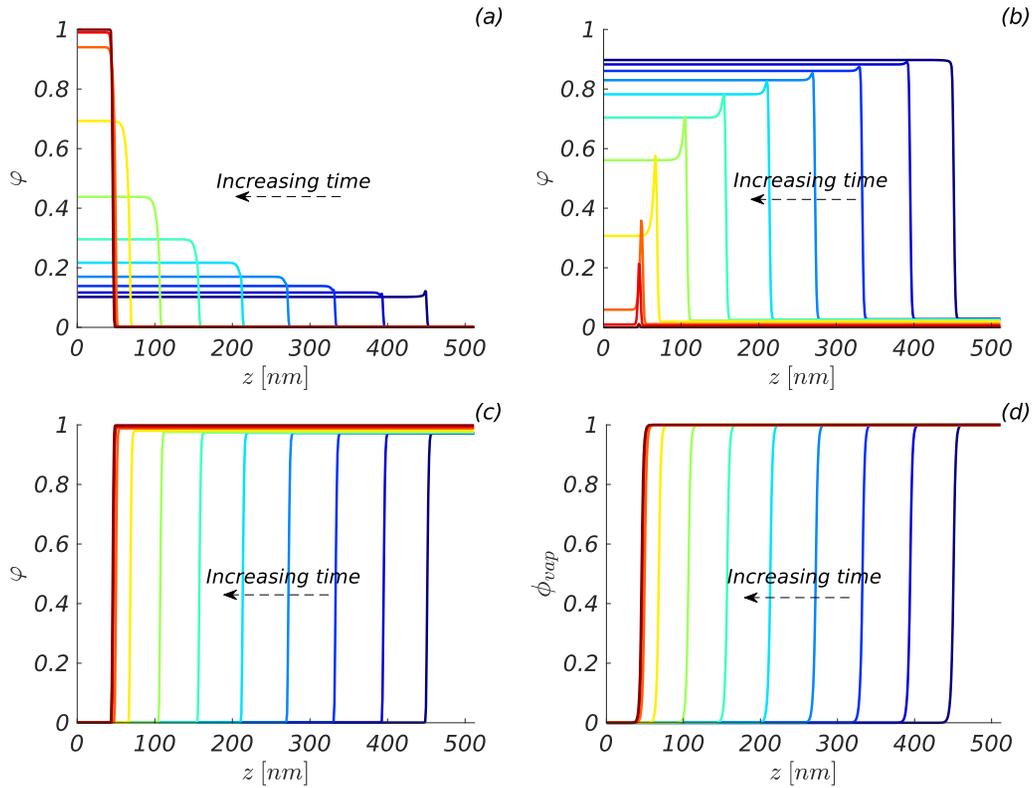

***Figure S15.*** *Typical volume fraction field for the solute (a), the solvent (b), the air (c) and the order parameter field (d) during drying. The interface moves from right to left from the initial state (blue) to the final state (brown). The system is discretized by 512 points. Note that the solvent peak is broader but not higher than for the bigger system (Figure 10 of the main text).*

## S4. Diffusion coefficients and skin effect for the PS-toluene mixture

The self-diffusion coefficients of polystyrene and toluene in the mixture are known to be strongly composition-dependent. Various data are available from the literature for the self-diffusion coefficients of toluene [2][3][4] and of polystyrene [5][6] at different concentrations, as well as for the mutual diffusion coefficient [34][30][31][7][8][9][10][11]. Nevertheless, we are not aware of all data being available for this molecular weight at room temperature, and we can only estimate the diffusion properties from the published data that are available for polymers of various molecular weight and polydispersity. We use the data from the paper of Zettl (2009) available for a molecular weight of 67kg/mol (polydispersity index 1.05) at room temperature. The data are available for low polymer volume fraction. We also use the data from the papers of Mueller (2012 and 2013) for diffusion data at high polymer concentration and 30°C. There, the molecular weights are 240kg/mol (polydispersity index 5.65) and 380kg/mol (polydispersity index 1.01), respectively. In addition, we use the value of $2 \cdot 10^{-9}$ m²/s for the diffusion coefficient of pure toluene.

We approximate the self-diffusion coefficients with simple mathematical functions:

- for the polymer self-diffusion coefficient, we use a simple power law, also known as Vignes law $D_{s,poly}(\varphi) = \prod_{k=1}^{n}\left(D_{s,poly}^{\varphi_k \rightarrow 1}\right)^{\varphi_k}$, where $D_{s,poly}^{\varphi_k \rightarrow 1}$ is the self-diffusion coefficient of the polymer in the $k$ pure materials. Note that this expression does not have the least physical meaning for polymers in solution, but it is very simple and on this particular case, the agreement with the data is quite good. Note that we do not expect the agreement to be good in the general case and one could think of modeling the diffusion coefficient of the polymer using the classical scaling laws from the reptation theory [12]

- for the solvent self-diffusion coefficient, we use the functional form proposed by Siebel [13] inspired by the free volume theory [14],

$$D_{s,solv}(\varphi) = exp\left(-\frac{a + b\sum_{k\,solvents}\varphi_k}{1 + c\sum_{k\,solvents}\varphi_k}\right) \tag{S40}$$

where $a = -ln\left(D_{s,solv}^{\varphi_{poly} \rightarrow 1}\right)$, $b = ln\left(D_{s,solv}^{\varphi_{poly}\rightarrow 1}\right) - (1 + c)ln(D_{s,solv}^{\varphi_{solv}\rightarrow 1})$ and $c$ is an adjustable parameter.

The Onsager coefficient $\Lambda_{11}$ is calculated with the slow mode theory (Equation (15) in the main text) and the mutual diffusion coefficient taking into account the thermodynamic factor (Equation S5 of the Supplementary information). The result of this procedure is shown in **Figure S16**.

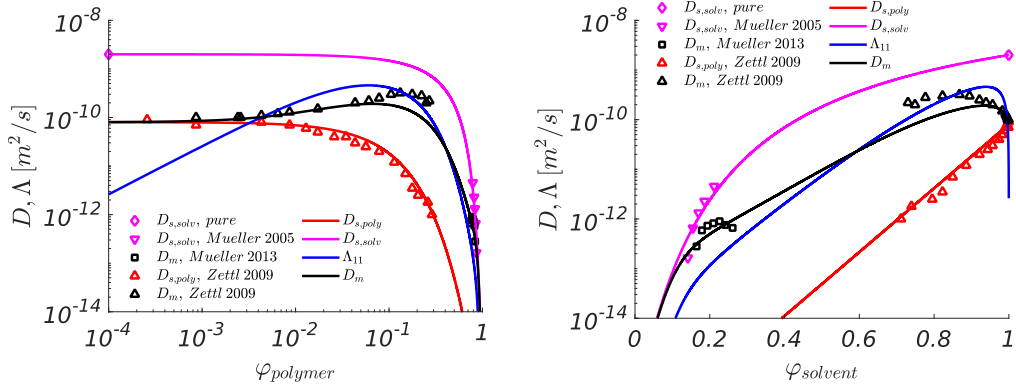

**Figure S16.** *Diffusion properties of the polystyrene-toluene mixture at different compositions ($D_{s,solv}$: self-diffusion coefficient of the solvent, $D_{s,poly}$: self-diffusion coefficient of the polymer, $D_m$: mutual diffusion coefficient, $\Lambda_{11}$: Onsage coefficient). Symbols are experimental values and lines calculated values.*

*The parameter used are $D_{s,solv}^{\varphi_{solv}\rightarrow 1} = 2 \cdot 10^{-9} m^2/s$, $D_{s,poly}^{\varphi_{solv}\rightarrow 1} = 8 \cdot 10^{-11} m^2/s$, $D_{s,solv}^{\varphi_{poly}\rightarrow 1} = 10^{-16} m^2/s$, $D_{s,poly}^{\varphi_{poly}\rightarrow 1} = 3 \cdot 10^{-17} m^2/s$, $c = 5$, $\chi = 0.35$ and $N_{poly}/N_{solv} = 327$*

*(Left) depending on the volume fraction of polymer, log-scale (Right) depending on the volume fraction of solvent, linear scale*

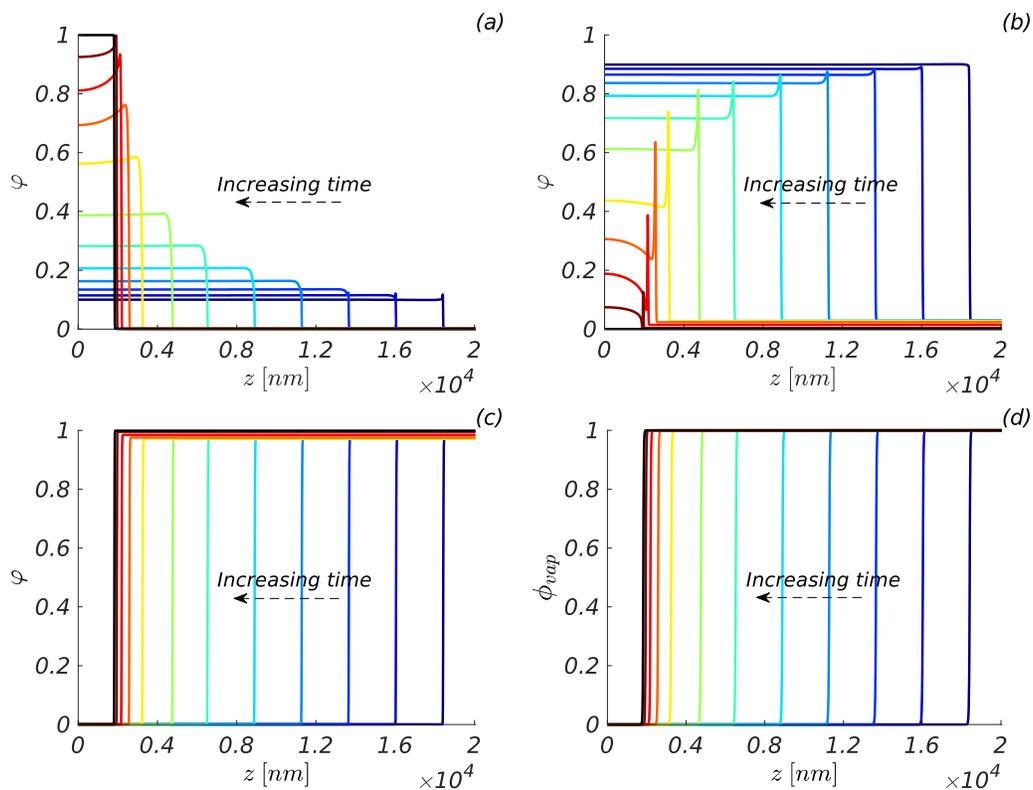

***Figure S17.*** *Typical volume fraction field for the solute (a), the solvent (b), the air (c) and the order parameter field (d) during drying of the polystyrene-toluene mixture. The interface moves from right to left from the initial state (blue) to the final state (black). The system is discretized by 2048 points.*

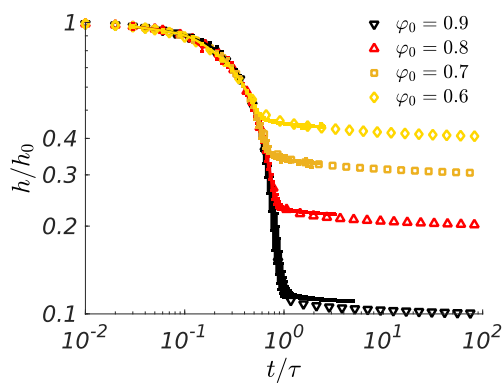

***Figure S18.*** *Simulated time dependent film height (symbols) for different volume fractions $\varphi_0$ of solvent in the initial film, and compared to the experimental results (full lines, reproduced from Ref. [15] with permission from the PCCP Owner Societies) for a 35kg/mol polystyrene – toluene mixture. Error bars represent +/- 1 standard deviation. $h_0$ is the initial height and $\tau$ the time needed to dry a film with $\varphi_0 = 0.9$ at constant drying rate. This is the same results as in Figure 13 of the main text, but with a log-log scale to visualize the long time behavior.*

# S5. Parameters used for the 2D simulations

| | |
|---|---|
| $T$ | 300 K |
| $\rho_i$ (all) | 1000 kg/m$^3$ |
| $v_0$ | 3·10$^{-5}$ m$^3$/mol |
| $N_{air}$ | 1 |
| $N_i$ (solute + solvent) | 3 |
| $\chi$ | 1.5 |
| $P_{sat,air}$ | 10$^8$ Pa |
| $P_{sat,solute}$ | 10$^2$ Pa |
| $P_{sat,solvent}$ | 3·10$^3$ Pa |
| $P_i^\infty$ | 0 Pa |
| $\kappa_i$ (all) | 15·10$^{-10}$ J/m |
| $\varepsilon_{vap}$ | 2·10$^{-4}$ (J/m)$^{0.5}$ |
| $D_{s,i}^{liq}$ (all) | 2·10$^{-9}$ m$^2$/s |
| $D_i^{vap}$ (all) | 10$^{-5}$ m$^2$/s |
| $M_{vap}$ | 10$^6$ s$^{-1}$ |
| $\alpha$ | -2.3·10$^{-5}$ |
| $\beta$ | 10$^{-5}$ |
| $\gamma$ | 1 |

Table S2. Parameter set for the 2D simulations of film drying on a rough substrate

# Literature

(1) Liu, X.; Vlugt, T. J. H.; Bardow, A. Predictive Darken Equation for Maxwell-Stefan Diffusivities in Multicomponent Mixtures. *Industrial & Engineering Chemistry Research* **2011**, *50* (17), 10350–10358. https://doi.org/10.1021/ie201008a.

(2) Mueller, F.; Naeem, S.; Sadowski, G. Toluene Sorption in Poly(Styrene) and Poly(Vinyl Acetate) near the Glass Transition. *Ind. Eng. Chem. Res.* **2013**, *52* (26), 8917–8927. https://doi.org/10.1021/ie302322t.

(3) Waggoner, R. A.; Blum, F. D.; MacElroy, J. M. D. Dependence of the Solvent Diffusion Coefficient on Concentration in Polymer Solutions. *Macromolecules* **1993**, *26* (25), 6841–6848. https://doi.org/10.1021/ma00077a021.

(4) Zielinski, J. M.; Duda, J. L. Predicting Polymer/Solvent Diffusion Coefficients Using Free-Volume Theory. *AIChE J.* **1992**, *38* (3), 405–415. https://doi.org/10.1002/aic.690380309.

(5) Zettl, U.; Hoffmann, S. T.; Koberling, F.; Krausch, G.; Enderlein, J.; Harnau, L.; Ballauff, M. Self-Diffusion and Cooperative Diffusion in Semidilute Polymer Solutions As Measured by Fluorescence Correlation Spectroscopy. *Macromolecules* **2009**, *42* (24), 9537–9547. https://doi.org/10.1021/ma901404g.

(6) Liu, R.; Gao, X.; Adams, J.; Oppermann, W. A Fluorescence Correlation Spectroscopy Study on the Self-Diffusion of Polystyrene Chains in Dilute and Semidilute Solution. *Macromolecules* **2005**, *38* (21), 8845–8849. https://doi.org/10.1021/ma0511090.

(7) Reis, R. A.; Nobrega, R.; Oliveira, J. V.; Tavares, F. W. Self- and Mutual Diffusion Coefficient Equation for Pure Fluids, Liquid Mixtures and Polymeric Solutions. *Chemical Engineering Science* **2005**, *60* (16), 4581–4592. https://doi.org/10.1016/j.ces.2005.03.018.

(8) Rauch, J.; Köhler, W. Collective and Thermal Diffusion in Dilute, Semidilute, and Concentrated Solutions of Polystyrene in Toluene. *The Journal of Chemical Physics* **2003**, *119* (22), 11977–11988. https://doi.org/10.1063/1.1623745.

(9) Krüger, K.-M.; Sadowski, G. Fickian and Non-Fickian Sorption Kinetics of Toluene in Glassy Polystyrene. *Macromolecules* **2005**, *38* (20), 8408–8417. https://doi.org/10.1021/ma050353o.

(10) Mueller, F.; Krueger, K.-M.; Sadowski, G. Non-Fickian Diffusion of Toluene in Polystyrene in the Vicinity of the Glass-Transition Temperature. *Macromolecules* **2012**, *45* (2), 926–932. https://doi.org/10.1021/ma202283e.

(11) Zielinski, J. M. A Friction Factor Analysis of the Coupling between Polymer/Solvent Self- and Mutual-diffusion: Polystyrene/Toluene. *Journal of Polymer Science: Part B: Polymer Physic* **1996**, *34*, 2759–2766. https://doi.org/10.1002/(sici)1099-0488(19961130)34:163.0.co.

(12) Rubinstein, M.; Colby, R. H. *Polymer Physics*; Oxford University Press: Oxford, New York, 2003.

(13) Siebel, D.; Scharfer, P.; Schabel, W. Prediction of Diffusion in a Ternary Solvent–Solvent–Polymer Blend by Means of Binary Diffusion Data: Comparison of Experimental Data and Simulative Results. *Journal of Applied Polymer Science* **2016**, *133* (36), 43899. https://doi.org/10.1002/app.43899.

(14) Vrentas, J. S.; Duda, J. L. Diffusion in Polymer—Solvent Systems. I. Reexamination of the Free-Volume Theory. *J. Polym. Sci. Polym. Phys. Ed.* **1977**, *15* (3), 403–416. https://doi.org/10.1002/pol.1977.180150302.

(15) Ronsin, O. J. J.; Jang, D.; Egelhaaf, H.-J.; Brabec, C. J.; Harting, J. A Phase-Field Model for the Evaporation of Thin Film Mixtures. *Phys. Chem. Chem. Phys.* **2020**, *22* (12), 6638–6652. https://doi.org/10.1039/D0CP00214C.